\documentclass[twocolumn,showpacs,
amsmath,amssymb,superscriptaddress]{revtex4-2}

\usepackage{graphicx}
\usepackage{dcolumn}
\usepackage{bm}
\usepackage[version=3]{mhchem} 
\usepackage{amsmath}
\usepackage{esvect}

\usepackage{array}
\usepackage{graphicx}
\usepackage{graphicx}
\usepackage{dcolumn}
\usepackage{bm}
\usepackage{color}
\usepackage{pifont}
\usepackage{natbib}
\usepackage{hyperref}
\hypersetup{
colorlinks = true,
urlcolor   = blue,
linkcolor  = blue,
citecolor  = blue
}
\usepackage{nicefrac}

\usepackage{color}
\newcommand{\red}[1]{{\color{black} #1}}

\begin{document}

\preprint{Preprint}

\title{Reproducing the asymptotic behavior of galaxy rotation curves by a novel constraint in general relativity}

\author{Godwill Mbiti Kanyolo}
\email{gmkanyolo@mail.uec.jp}
\affiliation{The University of Electro-Communications, Department of Engineering Science,\\ 
1-5-1 Chofugaoka, Chofu, Tokyo 182-8585}

\author{Titus Masese}
\email{titus.masese@aist.go.jp}
\affiliation{Research Institute of Electrochemical Energy (RIECEN), National Institute of Advanced Industrial Science and Technology (AIST), 1-8-31 Midorigaoka, Ikeda, Osaka 563-8577, Japan}
\affiliation{AIST-Kyoto University Chemical Energy Materials Open Innovation Laboratory (ChEM-OIL), Yoshidahonmachi, Sakyo-ku, Kyoto-shi 606-8501, Japan}


\begin{abstract}
The explanation for the non-Keplerian behavior of galaxy rotation curves in the weak gravity regime has been the center of scientific interest in cosmology for decades. The cold dark matter ($\rm CDM$) paradigm has been posited as the standard explanation, where for galaxies satisfying the Tully-Fisher relation, the mass of the dark matter halo from a large class of universal dark matter profiles ought to roughly increase linearly with radial distance at large distances, $m(r) \sim r/nG$ ($G$ is the gravitational constant and $n$ is a dimensionless parameter which depends on the amount of baryonic matter $M$ within the galaxy). Despite numerous theoretical and computational advances in modeling galaxy formation and evolution within the Lambda-cold dark matter ($\Lambda{\rm CDM}$) model in cosmology, a scientific consensus on the origin of the observed dependence of the dimensionless parameter $n = (GMa_{0})^{-1/2}$ on the mass of baryonic matter $M$ within the galaxy (the Tully-Fisher relation), and the connection of the cosmological constant $\Lambda$ to the parameter $a_{0} \sim (\Lambda/3)^{1/2}$ remains elusive. Here, we show that Einstein Field Equations with specific stress-energy-momentum sources can be remolded into $\nabla_{\nu}\mathcal{K}^{\nu}_{\,\,\mu} = 8\pi GM\Psi^{*}\mathcal{D}_{\mu}\Psi$, where $\mathcal{K}_{\mu\nu}$ is a complex Hermitian tensor, $\mathcal{D}_{\mu}$ is a covariant derivative and $\Psi$ is a complex-valued function. This avails a novel constraint, $\nabla_{\mu}\nabla_{\nu}\mathcal{K}^{\mu\nu} = 0$ not necessarily available in Einstein's General Relativity. We test the approach in the weak-gravity regime, where we can readily reproduce the Tully-Fisher relation using the usual charge-less pressure-less fluid as the source of the Einstein Field Equations, $G_{\mu\nu} + \Lambda g_{\mu\nu} = 8\pi GM\rho u^{\mu}u^{\nu}$ where $G_{\mu\nu}$ is the Einstein tensor, representing the cold dark matter. Moreover, the proposed approach in the weak-field regime is equivalent to a Ginzburg-Landau theory of $n$ bosons, where the order parameter, $\Psi = \sqrt{\rho}\exp(iS)$ with $S$ the action is normalized as $\int_{0}^{1/a_{0}} dr\,4\pi r^2\Psi^*\Psi = n$ and $1/a_{0} \sim (\Lambda/3)^{-1/2}$ is the cut-off length scale comparable to the size of the de Sitter universe. Our investigations present a new perspective by providing a framework that reproduces the mass-asymptotic speed relation in galaxies within the cold dark matter paradigm.

\end{abstract}

\maketitle


\section{Introduction}

Based on the baryonic matter observed in galaxies, Einstein's theory of general relativity 
coupled with the standard model of particle physics predicts that galaxy rotation curves ought to fall off with radial distance in a Keplerian fashion. However, numerous observations show instead \red{a non-Keplerian behavior for galaxies and galaxy clusters, which} historically led to the postulate of the dark matter paradigm.\cite{famaey2012modified, zwicky1937masses, babcock1939rotation, freeman1970disks, rubin1970rotation, rogstad1972gross, corbelli2000extended, faber1976velocity} Although this paradigm is presently supported by other observations consistent with the Lambda-cold dark matter ($\Lambda{\rm CDM}$) model in cosmology,\cite{trimble1987existence, bertone2005particle, copi1995big, clowe2006direct, aguirre2001problems, van2018galaxy, van2016high} \red{it faces some daunting challenges in reproducing certain trends and features of rotation curves in galaxies.} In particular, simulations with general relativity and the data from Cosmic Microwave Background (CMB) anisotropy (WMAP data)\cite{carroll2017introduction} reveal a rich energy composition of the universe, namely non-relativistic/cold matter ($\Omega_{\rm M} = 0.27 \pm 0.04$), relativistic matter/radiation ($\Omega_{\rm rad.} = 8.24 \times 10^{-5}$) and dark energy ($\Omega_{\Lambda} = 0.73 \pm 0.04$). This cosmological model (\red{$\Lambda$CDM}) requires that about 83\% of the non-relativistic matter be dark matter, which is \red{cold (pressure-less)} and has a negligible interaction cross-section\cite{huo2020structure, essig2019constraining} with itself and the rest of the baryonic matter and radiation \red{(uncharged)}. \red{However, the $\Lambda$CDM model faces challenges in conclusively accounting for the asymptotic behavior of galaxy rotation curves, which predict a stringent connection between the amount of baryonic matter within the galaxy to the amount of \red{cold} dark matter in the halo.\cite{mcgaugh2000baryonic}}

\red{A clear embodiment of these challenges is the baryonic Tully-Fisher relation, involving baryonic matter (stars and interstellar gas) empirically obtained by McGaugh and Stacy (2012), which necessitates cold dark matter simulation models to consider feedback processes between the cold dark matter and the baryonic matter in order to reproduce the incredibly small scatter obtained in the empirical data.\cite{eisenstein1997can} Since dark matter predominantly interacts with baryonic matter via gravity, it has been argued that it is unlikely that the incredibly small scatter in the baryonic Tully-Fisher relation would be as a result of feedback processes that regulate the partition of angular momentum between the cold dark matter and the baryonic matter\cite{eisenstein1997can}, requiring several parameters in simulations to account for. Moreover, it has been proposed that the difficulty, in part, lies in the assumption that cold-dark matter has a negligible cross-section with standard model particles and thus is dissipation-less.\cite{foot2015dissipative, ackerman2009dark} In fact, while some models consider the possibility of a fraction of the dark matter sector particles possesing an electric charge of about $10^{-6}g$, where $g$ is the electric charge\cite{de1990charged, munoz2018small}, others introduce a dark sector $U(1)$ gauge field which mediates interactions between standard model particles and particles in the dark sector.\cite{agrawal2017make, kamada2020unification, foot2015dissipative} Such models provide a channel for a fraction of the cold dark matter to dissipate energy forming oblate spheroid, disk or geoid mass density profiles which deviate from spherical symmetry.\cite{kramer2016updated, randall2014dark, fan2013double}} 

\red{Universal rotation curves under the conditions of spherical symmetry of the cold dark matter halo typically reproduce the singular isothermal profile proportional to $1/r^2$ at large radius \cite{keeton2001catalog, martel2003gravitational, persic1996universal}, \textit{albeit} with notable exceptions such as the Navarro-Frenk-White (NFW) halos, which are well-fitted by a $1/r^3$ profile at large radius.\cite{navarro1997universal} For galaxies satisfying the baryonic Tully-Fisher relation, the mass of cold dark matter halo ought to roughly increase linearly with radial distance at large distances, $m(r) \sim r/nG$, corresponding to a singular isothermal profile of the form, $1/4\pi Gnr^2$, where $G$ is the gravitational constant and $1/\sqrt{n} \propto \sigma_{\rm V}$ is proportional to the velocity dispersion, $\sigma_{\rm V}(M)$ which depends on the amount of baryonic matter, $M$ within the galaxy via the virial theorem. A} scientific consensus on the origin of the observed dependence of $n = (GMa_{0})^{-1/2}$ on the mass of baryonic matter $M$ within the galaxy \red{(the baryonic Tully-Fisher relation)}, and the connection of the cosmological constant $\Lambda$ to the parameter $a_{0} \sim (\Lambda/3)^{1/2}$ remains elusive.\cite{famaey2012modified, carroll2001cosmological, bekenstein2004relativistic, mcgaugh2015tale, mcgaugh2005balance, kroupa2012failures, clifton2012modified, rahvar2014observational, hossenfelder2019strong} \red{On the other hand, MOdified Newtonian Dynamics (MOND) accounts for the Tully-Fisher relation and related properties in galaxies without the need for dark matter.\cite{mcgaugh2000baryonic, mcgaugh2012baryonic} Despite this, a satisfactory relativistic counterpart to Einstein's General Relativity (referred to as MOdified gravity) that also replicates the successes of the $\Lambda \rm{CDM}$ model in galactic clusters and cosmological scales remains elusive.\cite{famaey2012modified}} 

The difficulty in formulating a successful approach to dark matter and/or MOdified gravity \red{that satisfactorily reproduces the trends and features of galaxies as well as in galaxy clusters and cosmological scales} lies in the stringent constraints in experiment and theory required in order to be simultaneously consistent with the standard model of particle physics and Einstein's General Relativity.\cite{sumner2002experimental, ishak2019testing, famaey2012modified, cvetivc2002m,cvetivc2005supersymmetric, cvetivc2005new} In fact, almost all mainstream theoretical treatments of dark matter and/or MOdified gravity require additional particle degrees of freedom and/or the postulation of new properties of space-time \red{not present in general relativity and/or the standard model of particle physics}.\cite{trimble1987existence, catena2014susy, foot2015dissipative, tamaki2008post, konno2008flat, sotani2010slowly, stabile2011rotation, stabile2013galaxy, diez2018horndeski} However, fairly recent approaches such as superfluid dark matter\cite{berezhiani2015theory, sharma2019equation} and entropic gravity\cite{verlinde2017emergent, verlinde2011origin} have garnered traction over traditional theories since they are moderately successful in combining the triumphs of MOND on galactic scales and the $\Lambda$CDM model on cosmological scales.\cite{brouwer2017first, hossenfelder2019strong}

In particular, this contemporary superfluid dark matter approach is essentially a Ginzburg-Landau (Gross-Pitaevskii) theory\cite{huebener2001ginzburg, gross1961structure, pitaevskii1961vortex} where the two-particle interaction term in the free energy function of the order parameter is replaced by a three-particle interaction term\footnote{A $|\Psi|^4$ term in the Ginzburg-Landau (Gross-Pitaevskii) free energy arises from a two particle interaction, whereas a $|\Psi|^6$ term arises from a three particle interaction. Superfluid dark matter theories correspond to a three-particle interaction term but with a negative sign\cite{berezhiani2015theory}}, which allows for a phase-transition of dark matter particles. Despite the growing popularity of this approach, the lack of a satisfactory relativistic counterpart to the theory severely impedes predictions on cosmological scales.\cite{hossenfelder2019strong} On the other hand, entropic gravity applies the holographic principle\cite{maldacena1999large, susskind1995world, bousso2002holographic} to the de Sitter universe, requiring that all effects attributed to dark matter merely arise from baryonic matter, when ideas on information theory and (quantum) gravity are taken into account.  

\red{In this paper, we propose relativistic equations of gravity given by, $\nabla_{\nu}\mathcal{K}^{\nu}_{\,\,\mu} = 8\pi GM\Psi^{*}\mathcal{D}_{\mu}\Psi$, where $\Psi$ is a complex-valued function, $\mathcal{D}_{\mu}$ a covariant derivative, $\mathcal{K}_{\mu\nu} = R_{\mu\nu} + i2g\mathcal{F}_{\mu\nu}$ is a complex Hermitian tensor comprising the Ricci tensor $R_{\mu\nu}$ and a novel gauge field strength $\mathcal{F}_{\mu\nu} = \partial_{\mu}\mathcal{A}_{\nu} - \partial_{\nu}\mathcal{A}_{\mu}$, $\mathcal{A}_{\mu}$ is the gauge field and $g$ is a coupling strength. This avails a novel constraint, $\nabla_{\mu}\nabla_{\nu}\mathcal{K}^{\mu\nu} = 0$ not necessarily available in Einstein's General Relativity. We test the approach in the weak-gravity regime, where we can readily reproduce the Tully-Fisher relation using the usual \red{uncharged} pressure-less fluid as the source of the Einstein Field Equations, $R_{\mu\nu} - \frac{1}{2}Rg_{\mu\nu} + \Lambda g_{\mu\nu} = 8\pi GM\rho u^{\mu}u^{\nu}$, where $R = R_{\mu\nu}g^{\mu\nu}$ is the Ricci scalar and $M\rho$ is the energy density of cold dark matter in the co-moving frame. Note that the stress-energy-momentum tensor of the gauge field does not couple to the Einstein Field Equations since the gauge field can be related to a Killing vector, $\xi_{\mu} \propto \mathcal{A}_{\mu}$ on any manifold with at least one space-time isometry, \textit{i.e.} $\mathcal{F}_{\mu\nu} \propto \nabla_{\nu}\xi_{\mu}$. Moreover, the proposed approach in the weak-field regime is equivalent to a Ginzburg-Landau theory} of $n$ bosons\cite{huebener2001ginzburg, gross1961structure, pitaevskii1961vortex}, where the order parameter\red{, $\Psi = \sqrt{\rho}\exp(iS)$ with $S$ the action,} is normalized as $\int_{0}^{1/a_{0}} dr\,4\pi r^2\Psi^*\Psi = n$ and $1/a_{0} \sim (\Lambda/3)^{-1/2}$ is the cut-off \red{length scale} comparable to the size of the de Sitter universe.\cite{hawking1973large} Our investigations present a new perspective 
by providing a framework that \red{reproduces the mass-asymptotic speed relation in galaxies within the cold dark matter paradigm.} 

Throughout this paper, we use Einstein's summation convention, and natural units where Planck's constant $\hbar$ and speed of light in vacuum $c$ are set to unity ($\hbar = c = 1$). We also assume a torsion-free $3 + 1$ dimensional space-time where $_{;\,\mu} \equiv \nabla_{\mu}$ is the metric compatible covariant derivative, \textit{i.e.} $\nabla_{\sigma}g_{\mu\nu} = g_{\mu\nu;\sigma} = 0$. \red{In section \ref{sec: SM}, we first introduce the field equations in the context of gauge symmetries, then in section \ref{sec: DM}, we consider a specific set of field equations in the context of dark matter. To aid the reader follow our arguments, we explicitly use script letters to distinguish standard model currents and fields, $J^{\mu}, A_{\mu}, D_{\mu}, F_{\mu\nu}, K_{\mu\nu}$ in section \ref{sec: SM} from the dark sector currents and fields, $\mathcal{J}^{\mu}, \mathcal{A}_{\mu}, \mathcal{D}_{\mu}, \mathcal{F}_{\mu\nu}, \mathcal{K}_{\mu\nu}$ in section \ref{sec: DM}.}

\red{\section{\label{sec: SM} Gravity and gauge symmetries}}

\red{\subsection{Trace of Einstein Field Equations}}

The relativistic equations of gravity are given by Einstein Field Equations\cite{einstein1997volume}, 
\begin{align}\label{EFE}
    R_{\mu\nu}-\frac{1}{2}Rg_{\mu\nu} + \Lambda g_{\mu\nu} = 8\pi G T_{\mu\nu},
\end{align}
where $R_{\mu\nu}$ is the Ricci tensor, $g_{\mu\nu}$ is the metric tensor, $R = R^{\mu\nu}g_{\mu\nu}$ is the Ricci scalar, $\Lambda$ is the cosmological constant, $T_{\mu\nu}$ is the stress-energy-momentum tensor and $G$ is the gravitational constant. It is straightforward to show that eq. (\ref{EFE}) satisfies the differential equations,
\begin{subequations}\label{DEFE}
\begin{align}\label{DEFEa}
    T^{\mu\nu}_{\,\,\,\,;\,\mu} = 0,\\
    R_{;\,\mu} = -8\pi GT_{;\,\mu},
    \label{DEFEb}
\end{align}
\end{subequations}
where $T = T^{\mu\nu}g_{\mu\nu}$ is the contracted stress-energy-momentum tensor. Note that eq. (\ref{DEFEa}) can be taken as a consequence of the equations of motion of matter and \red{gauge} fields in curved space-time, whereas eq. (\ref{DEFEb}) as the equations of motion of $g_{\mu\nu}$, which can be solved to yield $R = -8\pi G T + 4\Lambda$, the trace of the Einstein Field Equations.

\red{\subsection{Remolding Einstein Field Equations}}

Consider the stress-energy-momentum tensor of \red{charged particles and the electromagnetic (U($1$) gauge field) given by $T^{\mu\nu} = T^{\mu\nu}_{\rm particle} + T^{\mu\nu}_{\rm gauge}$ with,
\begin{subequations}
\begin{align}
    T^{\mu\nu}_{\rm particle} = -Tu^{\mu}u^{\nu},\\
    T^{\mu\nu}_{\rm U(1)} = F^{\mu\alpha}F^{\nu}_{\,\,\alpha} - \frac{1}{4}F^{\alpha\beta}F_{\alpha\beta}g^{\mu\nu},
\end{align}
\end{subequations}
the particle and gauge field} stress-energy-momentum tensors respectively\cite{tauber1979albert}, where $T = -M\rho$ is the particle mass density, $M$ is the mass of the particle, 
$u^{\mu} = dx^{\mu}/d\tau$ is the four-velocity of the particle satisfying $u^{\mu}u_{\mu} = -1$ with $x^{\mu}$ the space-time coordinates and $\tau$ an affine parameter, and $F_{\mu\nu} = \partial_{\mu}A_{\nu} -  \partial_{\nu}A_{\mu}$ is the radiation field strength with $A_{\mu}$ the electromagnetic U($1$) vector potential. Applying eq. (\ref{DEFEa}) to $T^{\mu\nu}$ yields the equations of motion for the gauge and particle respectively,
\begin{subequations}\label{MFE}
\begin{align}\label{MFEa}
    \nabla_{\mu}F^{\mu\nu} = g\rho u^{\mu} \equiv gJ^{\mu},\\
    Mu^{\nu}\nabla_{\nu}u^{\mu} = gu_{\nu}F^{\nu\mu},
    \label{MFEb}
\end{align}
where $g$ and $J^{\mu}$ are the U($1$) charge and current density respectively of the particle. Note that eq. (\ref{MFEb}) can be re-written as
\begin{align}\label{Action}
    S_{;\,\mu} = Mu_{\mu} + gA_{\mu},
\end{align}
\end{subequations}
where $S = -M\int d\tau + g\int A_{\mu}dx^{\mu}$ is the action of the particle of mass $M$ and charge $g$, by varying $S$ with respect to $x^{\mu}$. 

Introducing a complex-valued scalar field,
\red{
\begin{subequations}\label{Complex_Fields_eq}
\begin{align}
   \Psi = \sqrt{\rho}\exp iS, 
\end{align}
alongside a complex Hermitian tensor,
\begin{align}
    K_{\mu\nu} = R_{\mu\nu} + ig^* F_{\mu\nu},
\end{align}
\end{subequations}}
where $g^*$ is a constant with dimensions of charge to be determined, we can rewrite eq. (\ref{DEFE}) using eq. (\ref{MFE}) as complex (valued) field equations to yield,
\begin{align}\label{CFE}
    \nabla^{\nu}K_{\nu\mu} = 8\pi GM\Psi^{*}D_{\mu}\Psi, 
\end{align}
where $D_{\mu} = \nabla_{\mu} - igA_{\mu}$ is the U($1$) covariant derivative and,
\begin{align}\label{g_asterisk}
   g^* = 8\pi GM^2/g.
\end{align}
Making use of the Bianchi identity $\nabla^{\mu}R_{\mu\nu} = \frac{1}{2}\nabla_{\nu}R$, we can check that the imaginary and real parts of eq. (\ref{CFE}) simply yield eq. (\ref{DEFEa}), which are equivalent to eq. (\ref{MFE}) and (\ref{DEFEb}) respectively. 

A crucial question to pose is whether eq. (\ref{EFE}) and eq. (\ref{DEFE}) are exactly equivalent, particularly whether the set of all solutions of eq. (\ref{CFE})  labeled by the metric tensor $g'_{\mu\nu}$ is equivalent to the set of all solutions of the Einstein Field Equations given by eq. (\ref{EFE}), \textit{i.e.} is $g'_{\mu\nu} = g_{\mu\nu}$? Certainly, eq. (\ref{EFE}) satisfies the real part of eq. (\ref{CFE}) given by eq. (\ref{DEFEb}) with $4\Lambda$ the integration constant. However, the converse is not necessarily true unless the imaginary part of eq. (\ref{CFE}) guarantees eq. (\ref{DEFEa}) is also satisfied. In particular, since the trace-less energy-momentum tensor of electromagnetism does not couple to the Ricci scalar, $g'_{\mu\nu} = g_{\mu\nu}$ holds true \textit{if and only if} the imaginary part guarantees it. Moreover, since the U($1$) current $J^{\mu} = \rho u^{\mu}$ in eq. (\ref{MFEa}) is conserved, we are tempted to impose the additional constraint $\nabla^{\mu}(\Psi^{*}D_{\mu}\Psi) = \frac{1}{2}\nabla^{\mu}\nabla_{\mu}\rho + i\nabla^{\mu}(\rho u_{\mu}) = 0$. This would imply that $\nabla^{\mu}\nabla^{\nu}K_{\nu\mu} = 0$ and the Ricci scalar satisfies the massless real Klein-Gordon equation, $\nabla^{\mu}\nabla_{\mu}R = 0$. Thus, since this additional constraint is not necessarily guaranteed by eq. (\ref{EFE}), we conclude that $g'_{\mu\nu} \neq g_{\mu\nu}$ is also possible. 


\red{\subsection{Dirac and gauge fields}}

To incorporate the U($1$) Dirac current\cite{dirac1981principles} $\bar{\psi}\gamma_{\mu}\psi$ in eq. (\ref{Complex_Fields_eq}), we perform the \red{replacement,
\begin{subequations}
\begin{align*}
    \rho \rightarrow \bar{\psi}\psi,\\
    -M\int d\tau \rightarrow M\int\frac{\bar{\psi}\gamma_{\mu}\psi}{\bar{\psi}\psi}dx^{\mu}, 
\end{align*}
where $\psi$ is the Dirac spinor, $\bar{\psi} = \psi^{\dagger}(\gamma^0)^{-1}$ and $\gamma_{\mu}$ are the gamma matrices satisfying $\gamma_{\mu}\gamma_{\nu} + \gamma_{\nu}\gamma_{\mu} = 2g_{\mu\nu}$. Consequently, for all known U($1$) charged particles in the standard model, $g^*/g \sim 10^{-40} \ll 1$ due to the weakness of gravity relative to electromagnetism. A similar replacement can be carried out for SU($N$) Dirac currents\cite{yang1954conservation}, 
\begin{align*}
    \rho \rightarrow \bar{\psi}\psi,\\
    -M\int d\tau \rightarrow M\int \frac{(\bar{\psi}\gamma_{\mu}\tau_{a}\psi + f_{abc}A_{b\nu}F^{\,\nu}_{c\,\,\mu})\tau_{a}}{\bar{\psi}\psi}dx^{\mu},
\end{align*}
\end{subequations}}
where, $F_{\mu\nu} = (\partial_{\mu}A_{a\nu} - \partial_{\nu}A_{a\mu} + gf_{abc}A_{b\mu}A_{c\nu})\tau_{a} = \partial_{\mu}A_{\nu} - \partial_{\nu}A_{\mu} - ig[A_{\mu}, A_{\nu}]$ is the Yang-Mills curvature tensor\cite{yang1954conservation} where $A_{\mu} = A_{a\mu}\tau_{a}$ is the SU($N$) gauge field, $[\tau_{a}, \tau_{b}] = if_{abc}\tau_{c}$ is the SU($N$) algebra with $f_{abc}$ the structure constants and repeated roman/latin indices are summed over. The SU($N$) matrices are normalized as ${\rm Tr}(\tau_{a}\tau_{b}) = \delta_{ab}/2$. 

\red{These replacements follow from the stress-energy momentum tensors of the Dirac fields,
\begin{subequations}
\begin{multline}
    T^{\mu\nu}_{\rm Dirac} = \frac{1}{4i}\left ( \bar{\psi}\gamma^{\mu}D^{\nu}\psi + \bar{\psi}\gamma^{\nu}D^{\mu}\psi \right ) + \\
    \frac{1}{4i}\left ((\tilde{D}^{\mu}\bar{\psi})\gamma^{\nu}\psi + (\tilde{D}^{\nu}\bar{\psi})\gamma^{\mu}\psi
 \right ),
 \end{multline}
 the gauge field,
 \begin{align}
    T^{\mu\nu}_{\rm SU(N)} = 2{\rm Tr}\left ( F^{\mu\alpha}F^{\nu}_{\,\,\alpha} - \frac{1}{4}F^{\alpha\beta}F_{\alpha\beta}g^{\mu\nu}\right ),
\end{align}
and the Dirac equation in curved space time,
\begin{align}
    i\gamma^{\mu}(D_{\mu}\psi) = M\psi,\\
i(\tilde{D}_{\mu}\bar{\psi})\gamma^{\mu} = M\bar{\psi},
\end{align}
\end{subequations}
where $D_{\mu}\psi = (\partial_{\mu} - \frac{1}{4} \omega_{\mu}^{\bar{a}\bar{b}}\gamma_{\bar{a}}\gamma_{\bar{b}} - igA_{\mu})\psi$, $\tilde{D}_{\mu}\bar{\psi} = -\partial_{\mu}\bar{\psi} - \bar{\psi}(\frac{1}{4} \omega_{\mu}^{\bar{a}\bar{b}}\gamma_{\bar{a}}\gamma_{\bar{b}} + igA_{\mu})$, $\omega_{\mu}^{\bar{a}\bar{b}} = e^{\bar{a}}_{\,\,\alpha}\partial_{\mu}e^{\bar{b}\alpha} + \Gamma^{\beta}_{\,\,\mu\alpha}e^{\bar{a}}_{\,\,\beta}e^{\bar{b}\alpha}$ is the spin connection, $\Gamma^{\alpha}_{\,\,\mu\nu} = \frac{1}{2}g^{\alpha\beta}(\partial g_{\mu\beta}\partial x^{\nu} + \partial g_{\beta\nu}/\partial x^{\mu} - \partial g_{\mu\nu}/\partial x^{\beta})$ are the Christoffel symbols and $e_{\mu}^{\,\bar{a}}$ are the tetrad fields. 
}

\red{\subsection{The complex-valued function and the hierarchy problem}}

Of particular interest is the role of the charge $g^*$ and the complex function $\Psi$. It appears very unlikely that an unknown physical process would drive the coupling $g^*$ to such an infinitesimal value relative to $g$ coupling constants in the standard model of particle physics. Indeed, this is a variant of the well-known hierarchy problem in the standard model of particle physics.\cite{koren2020hierarchy} To make progress, recall that $g^*$ approaches $g$ in a Grand Unified Theory (GUT) scale\cite{buras1978aspects} where gravity and the gauge theory have a unified strength.\footnote{At this scale, the strong, weak and electromagnetic forces are governed by a simple lie group such as SU($N$), with a single coupling $g$. This is implicit here.} Moreover, eq. (\ref{CFE}) is invariant (\textit{i.e.} still reproduces the correct equations of motion) under the transformation \red{$A_{\mu} \rightarrow \epsilon^{-1} A_{\mu}$ and $u_{\mu} \rightarrow \epsilon^{-1} u_{\mu}$, where $\epsilon = g^*/2g = 4\pi GM^2/g^2$ is the relative strength between gravity and the gauge field. 

Thus, under this transformation, the complex field transforms as $\Psi \propto \exp (iS) \rightarrow \Psi' \propto \exp (iS') = \exp(iS/\epsilon)$. Since $S$ is the action of the particle, $\Psi'$ takes a form reminiscent of the wave function in Feynman's path integral formalism of quantum mechanics ($\psi \propto \exp(iS/\hbar)$, when $\hbar \neq 1$) and $\epsilon \rightarrow 0$ is akin to taking the classical limit $\hbar \rightarrow 0$ in quantum mechanics.\cite{brown2005feynman} This suggests that we should expect new physics (represented by $g_{\mu\nu} \rightarrow g'_{\mu\nu}$) described by $\Psi$ in eq. (\ref{CFE}) to be apparent in the inverse limit, $1/\epsilon \rightarrow 0$.

Unfortunately, due to the weakness of gravity relative to the other known fundamental forces ($\epsilon \sim 0$), the effects of $\Psi$ are restricted in favor of the Einstein Field Equations given by eq. (\ref{EFE}) in the present epoch.}

\red{\section{\label{sec: DM} Dark matter}}

\subsection{Novel U(1) field}

In the previous section, we remarked that the vanishing of the total divergence of eq. (\ref{CFE}) requires the Ricci scalar $R = 8\pi GM\rho + 4\Lambda$ to satisfy the massless Klein-Gordon equation \red{since} $\nabla^{\mu}\partial_{\mu}\rho = 0$ -- a condition not necessarily satisfied in general relativity. We shall implement this condition by introducing \textit{a priori} the following \textit{ansatzes}:
\begin{enumerate}
    \item  A conserved diffusion current given by,
    $D\partial_{\mu}\rho = \mathcal{J}_{\mu}$ where $D = 1/2M$ is the diffusion constant;
    \item A novel U(1) symmetry\red{, not present in the standard model of particle physics,} which guarantees that the diffusion current $\mathcal{J}^{\mu}$, couple\red{s} to the field strength $\mathcal{F}_{\mu\nu} = \partial_{\mu} \mathcal{A}_{\nu} - \partial_{\nu}\mathcal{A}_{\mu}$ as
    $\nabla_{\mu}\mathcal{F}^{\mu\nu} = \frac{1}{2}g^*\mathcal{J}^{\nu}$, where $g^* = 8\pi GM^2/g$ is given by eq. (\ref{g_asterisk}), is identically conserved. 
\end{enumerate}
Herein, we shall refer to $\mathcal{J}_{\mu} = D\partial_{\mu}\rho$ and the conservation law $\nabla_{\mu}\mathcal{J}^{\mu} = 0$ of the diffusion current as Fick's \textit{first} and \textit{second} laws respectively, due to their resemblance to the well-known diffusion laws.\cite{fick1855v} \red{Moreover, we explicitly use the script letters to distinguish standard model currents and fields, $J^{\mu}, A_{\mu}, D_{\mu}, F_{\mu\nu}, K_{\mu\nu}$ in the previous section from the dark sector currents and fields, $\mathcal{J}^{\mu}, \mathcal{A}_{\mu}, \mathcal{D}_{\mu}, \mathcal{F}_{\mu\nu}, \mathcal{K}_{\mu\nu}$.} 

Defining the gauge derivative $\mathcal{D}_{\mu} = \nabla_{\mu} - ig\mathcal{A}_{\mu}$, we can proceed to introduce the complex Hermitian tensor $\mathcal{K}_{\mu\nu} = R_{\mu\nu} + i2g\mathcal{F}_{\mu\nu}$ and the equations of motion, 
\begin{subequations}\label{QG}
\begin{align}
    \nabla^{\mu}\mathcal{K}_{\mu\nu} = 8\pi GM\Psi^*\mathcal{D}_{\nu}\Psi,
\end{align}
where $g^*$ fixes the argument of $\Psi = \sqrt{\rho}\exp(iS)$ to be given by, 
\begin{align}\label{action}
   S = M\int \xi_{\mu}dx^{\mu} + g\int \mathcal{A}_{\mu}dx^{\mu},
\end{align}
\end{subequations}
where we identify the four-velocity $\xi_{\mu}$ related to the conserved U($1$) current by $\mathcal{J}_{\mu} = D\partial_{\mu}\rho = \rho \xi_{\mu}$. 

\subsection{Thermodynamics}

Consider the Hamiltonian, $\mathcal{H}$ for a particle in a pressure-less fluid of number density $\rho$ given by, 
\begin{subequations}\label{Hamiltonian}
\begin{align}
\mathcal{H} = \frac{M}{2}g^{\mu\nu}\xi_{\mu}\xi_{\nu} + M\Phi, 
\end{align}
where $\Phi$ is a function of $x^{\mu}$ to be determined. The equations of motion for such a particle are given by the canonical equations, 
\begin{align}\label{canonical_1}
    M\frac{dx^{\mu}}{d\lambda} = \frac{\partial \mathcal{H}}{\partial \xi_{\mu}}, \,\,
    M\frac{d\xi_{\mu}}{d\lambda} = \frac{\partial \mathcal{H}}{\partial x^{\mu}}, 
\end{align}
which yield,
\begin{align}\label{geodesic}
    \frac{d^2x^{\sigma}}{d\lambda^2} + \Gamma^{\sigma}_{\,\,\mu\nu}\xi^{\mu}\xi^{\nu} = \xi^{\mu}\nabla_{\mu}\xi^{\sigma} = g^{\mu\sigma}\frac{\partial \Phi}{\partial x^{\mu}}, 
\end{align}
\end{subequations}
where $\lambda$ is an affine parameter, $dx^{\mu}/d\lambda = \xi^{\mu}$ and  $\Gamma^{\sigma}_{\,\,\mu\nu} = \frac{1}{2}g^{\sigma\rho}(\partial g_{\mu\rho}/\partial x^{\nu} + \partial g_{\nu\rho}/\partial x^{\mu} - \partial g_{\mu\nu}/\partial x^{\rho})$ are the Christoffel symbols. 

Introducing the Boltzmann factor by setting the number density to be equal to $\rho \propto \exp(-\bar{\beta} 2\mathcal{H})$, where the four-velocity $\bar{\beta}\nabla_{\mu}\Phi = \xi_{\mu}$ is also the drift velocity and $\bar{\beta} = \beta/2\pi = 1/2\pi k_{\rm B}T$ is the reduced inverse temperature, guarantees the diffusion current can also be written as $\mathcal{J}_{\mu} = \rho \xi_{\mu}$, as long as $g^{\mu\nu}\xi_{\mu}\xi_{\nu}$ is coordinate independent. The Boltzmann factor introduces the remaining fundamental constant (the Boltzmann constant, $k_{\rm B}$ alongside Planck's constant, $\hbar$, the gravitational constant, $G$ and the speed of light in vacuum, $c$), useful in defining the Planck scale.\cite{bingham2007can}

Recall that according to eq. (\ref{QG}), the action of such a particle is given by eq. (\ref{action}), which corresponds to the following equations of motion,
\begin{align}\label{geodesic_2}
    \xi^{\rho}\nabla_{\rho}\xi^{\sigma} = \frac{g}{M}\xi_{\rho}\mathcal{F}^{\rho\sigma}.
\end{align}
Comparing eq. (\ref{geodesic}) and eq. (\ref{geodesic_2}), we require that,
\begin{align}\label{friction}
    \bar{\beta}\partial_{\sigma}\Phi = \xi_{\sigma} = \frac{g}{M}\bar{\beta}\xi^{\rho}\mathcal{F}_{\rho\sigma} = 2g\mu\xi^{\rho}\mathcal{F}_{\rho\sigma}, 
\end{align}
leading to a formula for the mobility of the particles $\mu = \bar{\beta} D$ (Einstein-Smoluchoski relation) since, according to Fick's \textit{first} law, $D = 1/2M$ is the diffusion coefficient.\footnote{Confer eq. (\ref{friction}) with $\vec{v} = g\mu\vec{E}$, where $\vec{v}$ is the velocity and $\vec{E}$ is the electric field}

\subsection{Geodesics}

Any affine parameter, $\lambda$ is unique upto a transformation $\lambda = (d\lambda/d\tau)\tau + \lambda_0$ representing the equation of a straight line where $\tau$ is another affine parameter and $\lambda_{0}$ and $d\lambda/d\tau$ are constants corresponding to the $y$-intercept and gradient respectively.\footnote{This is just the definition of an affine parameter. This is the unique transformation of the parameter $\tau$ which leaves the geodesic equation invariant.} We proceed to parameterize the gradient as $d\lambda/d\tau = \exp(-\Phi)$ using the function $\Phi$, where $d\Phi/d\tau = 0$, since $d\lambda/d\tau$ is a constant function of $\tau$ (\textit{i.e.}, $d^2\lambda/d\tau^2 = 0$). The four-velocity can be defined as $u^{\mu} = dx^{\mu}/d\tau = (d\lambda/d\tau)dx^{\mu}/d\lambda \equiv \exp(-\Phi)\xi^{\mu}$. In the limit $\Phi \ll 1$ corresponding to $\exp(2\Phi) \simeq 1 + 2\Phi$, 
the Hamiltonian transforms into,
\begin{subequations}
\begin{align}\label{Hamiltonian_2}
    \mathcal{H} = \frac{M}{2}\exp(2\Phi)g^{\mu\nu}u_{\mu}u_{\nu} + M\Phi \simeq \frac{M}{2}g^{\mu\nu}u_{\mu}u_{\nu},
\end{align}
where we have used $g^{\mu\nu}u_{\mu}u_{\nu} = u^{\mu}u_{\mu} = -1$. An identical procedure as before, 
\begin{align}\label{canonical_2}
    M\frac{dx^{\mu}}{d\tau} = \frac{\partial \mathcal{H}}{\partial u_{\mu}}, \,\,
    M\frac{du_{\mu}}{d\tau} = \frac{\partial \mathcal{H}}{\partial x^{\mu}},
\end{align}
\end{subequations}
produces the equations of motion for the four-velocity $u^{\mu}$. Thus, plugging in eq. (\ref{Hamiltonian_2}) into eq. (\ref{canonical_2}) yields the geodesic equation, 
\begin{align}\label{geodesic_1}
    \frac{d^2x^{\sigma}}{d\tau^2} + \Gamma^{\sigma}_{\,\,\mu\nu}u^{\mu}u^{\nu} = u^{\rho}\nabla_{\rho}u^{\sigma} \simeq 0, 
\end{align}
where we have used $dx^{\mu}/d\tau = u^{\mu}$ and the Christoffel symbols $\Gamma^{\sigma}_{\,\,\mu\nu} = \frac{1}{2}g^{\sigma\rho}(\partial g_{\mu\rho}/\partial x^{\nu} + \partial g_{\nu\rho}/\partial x^{\mu} - \partial g_{\mu\nu}/\partial x^{\rho})$.

By virtue of the anti-symmetry of $\mathcal{F}_{\mu\nu}$ in eq. (\ref{friction}), the drift velocity $\xi^{\mu}$ satisfies $g^{\mu\nu}\xi_{\mu}\xi_{\nu} = 0$, thus requiring that the first term of the Hamiltonian, $\mathcal{H}$ in eq. (\ref{Hamiltonian}) to identically vanish. 
Consequently, the conservation of the U($1$) current, $\nabla_{\mu}(\rho \xi^{\mu}) = 0$ equivalent to $\nabla_{\mu}\xi^{\mu} = -\frac{d}{d\lambda}\ln (\rho/\rho_{\rm c})$ yields $\nabla_{\mu}\xi^{\mu} = 2M\bar{\beta}\frac{d}{d\lambda}\Phi = M\bar{\beta} \xi^{\mu}\nabla_{\mu}\Phi = 2M\xi^{\mu}\xi_{\mu} = 0$. We thus conclude that the potential $\Phi$ is a conserved quantity along the trajectory parameterized by $\lambda$, and ascribe this observation to the presence of the U($1$) symmetry introduced by the gauge field, $\mathcal{A}_{\mu}$. 

Additionally, using the continuity equation given by $\nabla_{\mu}(\rho u^{\mu}) = 0$ equivalent to $\nabla_{\mu}u^{\mu} = -\frac{d}{d\tau}\ln (\rho/\rho_{\rm c})$, we can solve for the number density to yield $\rho = \rho_{\rm c}\exp(-\int d\tau \nabla_{\mu}u^{\mu})$, where $\rho_{\rm c}$ is a proportionality constant playing the role of critical density. 
The Boltzmann factor corresponds to $u^{\mu} = \exp(2\bar{\beta} M\Phi)U^{\mu}$, where $U^{\mu}$ is a four-vector satisfying $\nabla_{\mu}U^{\mu} = 0$. This is straightforward to see since the exponent becomes, $-\int d\tau \nabla_{\mu}u^{\mu} = -2\bar{\beta} M\int d\tau \exp(2\bar{\beta} M\Phi)U^{\mu}\nabla_{\mu}\Phi = -2\bar{\beta} M\int d\tau\, u^{\mu}\nabla_{\mu}\Phi = -2\bar{\beta} M\Phi$, where we have used $u^{\mu}\nabla_{\mu}\Phi = d\Phi/d\tau$. As a result, the Boltzmann factor mimics successive affine transformations, 
\begin{subequations}
\begin{align}\label{n_times}
    u^{\mu} = \frac{d\lambda_1}{d\tau}\frac{d\lambda_2}{d\lambda_1}\cdots \frac{d\lambda_{2n}}{d\tau}\frac{dx^{\mu}}{d\lambda_{2n}} = \exp(2n\Phi)U^{\mu},
\end{align}
where $2n$ is the number of independent transformations and,
\begin{align}\label{equilibrium}
    n = \bar{\beta} M.
\end{align}
Thus, Fick's \textit{second} law merely guarantees the equilibrium condition provided the condition in eq. (\ref{equilibrium}) is satisfied. 

Finally, \red{eq. (\ref{geodesic_1})} motivates the equation of state to be given by,
\begin{align}\label{eq_of_state}
    G_{\mu\nu} + \Lambda g_{\mu\nu} \simeq 8\pi GM \rho u_{\mu}u_{\nu},
\end{align}
\end{subequations}
where $G_{\mu\nu} = R_{\mu\nu} - \frac{1}{2}Rg_{\mu\nu}$ is the Einstein tensor and the Bianchi identity ($\nabla_{\mu}G^{\mu\nu} = 0$) guarantees the geodesic equation and simultaneously the equilibrium condition are satisfied. Note that the trace of the equation of state is the solution of the real part of eq. (\ref{QG}), as expected. \red{This equation is surprising, since the stress-energy-momentum of the novel U($1$) does not appear on the right-hand side even though $\Psi$ couples to $\mathcal{F}_{\mu\nu}$ via the charge $g^*$. This serendipitous result is understood in section \ref{sec: Killing} by taking $\mathcal{A}_{\mu} \propto \xi_{\mu}$ and treating $\xi_{\mu}$ as a Killing vector. Since Killing vectors represent space-time isometries, functions of $\mathcal{F}_{\mu\nu} \propto \nabla_{\mu}\xi_{\nu}$ should not appear as sources in the Einstein Field Equations. Thus, eq. (\ref{eq_of_state}) is physically well-motivated since the right-hand side corresponds to the stress-energy-momentum tensor for an \red{uncharged} pressure-less ($P = 0$) perfect fluid given by $T^{\mu\nu}_{\rm fluid} = (M\rho + P)u^{\mu}u^{\mu} + Pg^{\mu\nu}$ corresponding to the stress-energy momentum tensor of cold dark matter.}

\subsection{\label{sec: Killing} Killing vectors and thermal equilibrium}

Conservation laws in general relativity arise from the presence of space-time isometries on the manifold. Thus, a further analysis of the geodesics above ought to be carried out by considering a solution for eq. (\ref{geodesic_1}) involving infinitesimal generators of isometries on the manifold (Killing vectors). Using eq. (\ref{action}), and setting $\partial S/\partial x^{\mu} = -g\mathcal{A}_{\mu}$ results in $\mathcal{A}_{\mu} = -M\xi_{\mu}/2g$, \red{which in turn leads to,} 
\begin{subequations}\label{horizon}
\red{
\begin{align}
   \mathcal{F}_{\mu\nu} = -\frac{M}{g}\nabla_{\mu}\xi_{\nu}, 
\end{align}
}
and hence transforms eq. (\ref{friction}) into, 
\begin{align}
    \xi^{\rho}\nabla_{\rho}\xi^{\mu} = \kappa\xi^{\mu},
\end{align}
\end{subequations}
where $\kappa = 1/\bar{\beta}$ is the surface gravity of the Killing horizon ($\xi^{\mu}\xi_{\mu} = 0$) when $\xi_{\mu}$ is a Killing vector satisfying $\nabla_{\mu} \xi_{\nu} = -\nabla_{\nu} \xi_{\mu}$.

In particular, introducing a time-like Killing vector $\xi^{\mu} = (1, \vec{0})$ has the advantage of guaranteeing thermal equilibrium on the Boltzmann factor, $\xi^{\mu}\nabla_{\mu}\rho = -2\bar{\beta} M \rho \xi^{\mu}\nabla_{\mu}\Phi = 2\bar{\beta} M \rho \partial\Phi/\partial t = 0$ since the Lie derivative of the Ricci scalar in the direction of a Killing vector identically vanishes. This is straightforward to prove using the Bianchi identity $\nabla_{\mu}R^{\mu\nu} = \frac{1}{2}g^{\nu\mu}\nabla_{\mu}R$ and using the well-known identity $\nabla_{\mu}\nabla_{\nu}\xi_{\sigma} = R^{\rho}_{\,\,\mu\nu\sigma}\xi_{\rho}$ for Killing vectors, where $R^{\rho}_{\,\,\mu\nu\sigma} = \partial_{\nu}\Gamma^{\rho}_{\,\,\mu\sigma} - \partial_{\sigma}\Gamma^{\rho}_{\,\,\mu\nu} + \Gamma^{\alpha}_{\nu\mu}\Gamma^{\rho}_{\,\,\alpha\sigma} - \Gamma^{\alpha}_{\sigma\mu}\Gamma^{\rho}_{\,\,\alpha\nu}$ is the Riemann curvature tensor, $R_{\mu\nu} = g^{\rho\sigma}R_{\rho\mu\sigma\nu}$ is the Ricci tensor and $g^{\mu\nu}R_{\mu\nu} = R$ is the Ricci scalar, which is related to the number density $\rho$ by $R = 8\pi GM\rho + 4\Lambda$. 

\red{However, typically we shall have a combination of two or more vectors, \textit{e.g.} $\xi^{\mu} = k^{\mu} + \Omega n^{\mu}$ where in a suitable coordinate system, $k^{\mu} = (1, \vec{0})$ and $n^{\mu} = (0, \vec{n})$ are time-like and space-like Killing vectors respectively and $\Omega$ is a constant that can be sent to zero ($\Omega \rightarrow 0$) to recover $\xi^{\mu} = (1, \vec{0})$ above. Finally, eq. (\ref{friction}) is solved by observing that, 
\begin{subequations}
\begin{align}
    \partial_{\alpha}\Phi = \frac{g}{M}\partial_{\alpha}(g^{\mu\nu}\xi_{\mu}\mathcal{A}_{\nu}) = -\frac{1}{2}\partial_{\alpha}(g^{\mu\nu}\xi_{\mu}\xi_{\nu}),
\end{align}
where we have used $\mathcal{A}_{\mu} = -M\xi_{\mu}/2g$ as before. This condition is equivalent to,
\begin{align}
    \partial_{\alpha}\mathcal{H} = 0,
\end{align}
where $\mathcal{H}$ is the Hamltonian given in eq. (\ref{Hamiltonian}).
\end{subequations}
}

\subsection{Poisson equation}

Moreover, we can identify eq. (\ref{geodesic}) as a Langevin equation\cite{lemons1997paul} with $g^{\mu\nu}\nabla_{\mu}\Phi = \kappa \xi^{\nu}$ the friction term and $\bar{\beta} = 1/\kappa$ taking the role of the mean-free time (average time between collisions), and eq. (\ref{eq_of_state}) as a Fokker-Planck equation\cite{platen1986risken}, where $\rho$ can be interpreted as the number density. 

To demonstrate this, we first consider the Newtonian limit of eq. (\ref{eq_of_state}) given by,
\begin{subequations}
\begin{align}\label{Poisson_1}
    R_{00} = \frac{\partial}{\partial x^\mu}\Gamma^{\mu}_{\,\,00} \simeq -\frac{1}{2}\nabla^2 g_{00} = 4\pi GM\rho - \Lambda, 
\end{align}
where $\xi^{\mu}\xi_{\mu} = u^{\mu}u_{\mu}\exp(2\Phi) = g_{00} \simeq -(1 + 2\Phi)$ and $\Phi \ll 1$. Here, $\nabla^2$ is the Laplacian, $u^{\mu} = (1, \vec{v})$ and $u_{\mu} = (g_{00}, \vec{v})$ where $\vec{v} \simeq 0$ is the co-moving condition, and $g_{\mu\nu}$ is diagonalized with the diagonal terms given by $\left \{ -(1 + 2\Phi), 1, 1, 1 \right \}$ corresponding to $\left \{ g_{00}, g_{11}, g_{22}, g_{33} \right \}$. Thus, plugging in $g_{00} \simeq -(1 + 2\Phi)$ into eq. (\ref{Poisson_1}) yields,
\begin{align}\label{Poisson_2}
    \nabla^2\Phi \simeq 4\pi GM\rho - \Lambda,
\end{align}
where $\Phi$ is the Newtonian potential. Meanwhile, setting $\xi^{\mu} = (1, -\vec{V})$ and hence $\xi_{\mu} = (g_{00}, -\vec{V})$ (where $\xi^{\mu}\xi_{\mu} = 0$ gives $g_{00} = -\vec{V}\cdot\vec{V}$) and plugging it into eq. (\ref{friction}) yields the mobility equation,
\begin{align}
    \bar{\beta}\partial \Phi/\partial t = g_{00} = 2g\mu\vec{V}\cdot\mathcal{\vec{E}},\\
    \bar{\beta}\vec{\nabla}\Phi = -\vec{V} = 2g\mu(\vec{\mathcal{E}} + \vec{V}\times\vec{\mathcal{B}}),
\label{Mobility_eq}
\end{align}
\end{subequations}
where $\mu = \bar{\beta} D$, $D = 1/2M$, $\mathcal{F}_{0i}$ and $\frac{1}{2}\varepsilon_{ijk}\mathcal{F}^{jk}$ are the components of the vectors $\vec{\mathcal{E}}$ and $\vec{\mathcal{B}}$ respectively which satisfy $\vec{\nabla}\cdot\vec{\mathcal{E}} = (g^*/2)\rho$ and $\vec{\nabla}\times\vec{\mathcal{B}} = (g^*/2)\rho\vec{V} + \partial \vec{\mathcal{E}}/\partial t$ by virtue of the \textit{a priori} \textit{ansatz} 2. The divergence of the expression in eq. (\ref{Mobility_eq}) merely reproduces eq. (\ref{Poisson_2}) with $\vec{V} \rightarrow 0$ and $\Lambda \rightarrow 0$.

Finally, it is apparent that the trial function, $\rho = \rho_{\rm c}\exp(2\varepsilon t - 2\bar{\beta} M\Phi)$, where $\varepsilon \rightarrow 0$ is a small parameter, plugged into eq. (\ref{Poisson_2}) can readily reproduce the Fokker-Planck equation, 
\begin{align}\label{Fokker-Planck}
    \frac{\partial \rho}{\partial t} = -\vec{\nabla}\cdot(\rho \vec{V}) + D\nabla^2\rho,
\end{align}
where the cosmological constant is given by $\Lambda = 2\varepsilon/\bar{\beta} \rightarrow 0$. This is an intriguing result since it not only justifies referring to conditions 1 and 2 above as Fick's laws, but also suggests a direct link of the geodesic equations and Einstein Field Equations to It\^{o} processes.\cite{ross1996stochastic} This link is the subject of further studies. 

\section{\label{sec: Results} Results}

\subsection{Asymptotic behavior of galaxy rotation curves}

\red{Universal rotation curves under the conditions of spherical symmetry of the cold dark matter halo typically reproduce the singular isothermal profile proportional to $1/r^2$ at large radius \cite{keeton2001catalog, keeton2001catalog, martel2003gravitational, persic1996universal}, \textit{albeit} with notable exceptions such as the Navarro-Frenk-White (NFW) halos, which are well-fitted by a $1/r^3$ profile at large radius\cite{navarro1997universal}. For galaxies satisfying the baryonic Tully-Fisher relation, the mass of cold dark matter halo ought to roughly increase linearly with radial distance at large distances, $m(r) \sim r/nG$, corresponding to a singular isothermal profile of the form, $1/4\pi Gnr^2$, where $G$ is the gravitational constant and $1/\sqrt{n} \propto \sigma_{\rm V}$ is proportional to the velocity dispersion, $\sigma_{\rm V}(M)$ which depends on the amount of baryonic matter, $M$ within the galaxy via the virial theorem. The} Poisson equation given in eq. (\ref{Poisson_2}) with $\rho = \rho_{\rm c}\exp(-2n\Phi(r)) \propto 1/r^2$ takes the singular isothermal form\cite{keeton2001catalog} typically used to model the density profiles of spherically symmetric dark matter hallows in spiral galaxies which satisfy the Tully-Fisher relation\cite{chan2013reconciliation}, since the solution for $\Phi(r)$ with $\Lambda = 0$ takes the form,
\begin{subequations}
\begin{align}\label{log_eq}
    \Phi(r) = \frac{1}{2n}\ln(Kr^2),
\end{align}
where $M$ is the mass of baryonic matter within the galaxy, $n = \bar{\beta} M$ and $K$ are constants to be determined ($K$ is determined by comparing eq. (\ref{density_profile}) and eq. (\ref{critical_2})). This can be confirmed by plugging the solution in eq. (\ref{log_eq}) back into the Poisson equation to yield,
\begin{multline}\label{Emden-Chandrasekhar_eq}
\nabla^{2}\Phi = \frac{1}{nr^2} = \frac{K}{n}\exp\left (-\ln (Kr^2) \right )\\ \simeq 4\pi GM\rho_{\rm c}\exp(-2n\Phi) = 4\pi GM\rho(r),
\end{multline}
\end{subequations}
where we determine the critical density and the density profile respectively to be given by,
\begin{align}\label{density_profile}
   \rho_{\rm c} = \frac{K}{4\pi GMn},\,\, \rho(r) = \frac{1}{4\pi GMnr^2}.
\end{align}
For the relativistic counterpart to eq. (\ref{Emden-Chandrasekhar_eq}), see eq. (\ref{space_space}). 

In 2 (spatial) dimensions, this Poisson equation is simply Liouville's equation with $K$ proportional to the Gaussian curvature of the 2 dimensional manifold.\cite{kanyolo2020idealised} However, this is not the case here since the halo is typically assumed to form a singular isothermal sphere profile in 3 dimensions, and hence it is referred to as the Emden-Chandrasekhar equation. Due to the singularity at $r \rightarrow 0$ of the density profile (eq. (\ref{density_profile})), this solution is typically assumed to only be valid at low acceleration and large $r$. Nonetheless, for our purposes herein, it will be sufficient to consider eq. (\ref{log_eq}) for large $r$. 

Moreover, in our approach $\rho = \Psi^*\Psi$ corresponds to a number density. The Tully-Fisher relation, which corresponds to the empirically obtained result for the flattening of galaxy rotation curves, is accurately reproduced under the normalization condition, 
\begin{subequations}
\begin{multline}\label{normalization}
\int_{\rm vol.} d^{\,3}x\, \Psi^*\Psi 
= \int_{\rm vol.} r^2drd\Omega \,\rho(r)\\
= \int_{0}^{r_{\infty} = 1/a_{0}}4\pi r^2 dr\,\rho(r)
= \frac{1}{GMa_0n} = n, 
\end{multline}
to yield, $1/n = (GMa_0)^{1/2}$ where $\int d\Omega = \int \sin\theta d\theta d\varphi = 4\pi$ is the total solid angle, $r$, $\theta$, $\varphi$ are the spherical (radial, colatitude and longitude) coordinates respectively and $r_{\infty} = 1/a_{0}$ is a radial cut-off size of the universe, introduced since the integral otherwise diverges. Thus, this cut-off scale can reasonably be taken to be comparable to the size of the de Sitter universe, $1/a_0 \sim \sqrt{3/\Lambda}$. Consequently, the amount of matter, $m(r)$ within a given radius $r$, needed to fit observations (the Tully-Fisher relation) can be calculated from, 
\begin{align}\label{dark_matter}
    m(r) = M\int_{0}^{r} 4\pi r'^2\rho(r')dr' = \frac{r}{nG}.
\end{align}
\end{subequations}
Correspondingly, the Newtonian limit of the geodesic equation in eq. (\ref{geodesic_1}) given by,
\begin{align}
    \frac{d^2x^{i}}{dt^2} \simeq -\Gamma_{00}^{i} \simeq \frac{1}{2}\frac{\partial g_{00}}{\partial x^{i}} \simeq -\frac{\partial \Phi}{\partial x^{i}},
\end{align}
which in spherical coordinates restricted to the plane of orbit ($t$, $r$, $\theta$ and $\varphi = 0$) transforms into,
\begin{subequations}
\begin{align}
    \frac{d^2r}{dt^2} - r\left ( \frac{d\theta}{dt} \right )^2 = -\frac{\partial}{\partial r}\Phi(r),\\
    \frac{1}{r}\frac{d}{dt}\left ( r^2\frac{d\theta}{dt} \right ) = 0,
\end{align}
\end{subequations}
gives the equations of motion for a star in orbit at the periphery of the galaxy at large $r$. The second equation is the conservation of angular momentum $L = mvr$, where $dr/dt = v = rd\theta/dt$ is the speed of the star in orbit. The star orbits at constant speed $d^2r/dt^2 = 0$ when $r(d\theta/dt)^2 = v_{\rm c}^2/r = \partial \Phi(r)/\partial r = 1/nr$, where we have plugged in the result of $\Phi(r)$ given in eq. (\ref{log_eq}). 

This readily yields the celebrated mass-asymptotic speed relation\cite{milgrom1983modification}, 
\begin{align}\label{asymptotic speed}
    v_{\rm c} = \frac{1}{n^{1/2}} = (GMa_0)^{1/4},
\end{align}
where we have used the result in eq. (\ref{normalization}) given by $1/n = (GMa_0)^{1/2}$. Finally, the condition given in eq. (\ref{equilibrium}) gives the temperature, $T = 1/\bar{\beta} k_{\rm B} 2\pi = M/2\pi nk_{\rm B} = M^{3/2}G^{1/2}a_0^{1/2}/k_{\rm B}$. 

\subsection{Phase-Transition}

It was remarked herein that the complex-valued function $\Psi$ has the property that $\Psi^*\Psi \neq 0$ represents the existence of dark matter, whose relevance is linked to considering the finite size of the universe. The question remains on the role of $\Psi$ in the field equations. This can be probed further by considering the Newtonian limit of eq. (\ref{QG}) given by the Poisson equation (eq. (\ref{Poisson_2})) and Fick's \textit{second} law corresponding to the real and imaginary parts of the field equations respectively. 

Plugging in the Boltzmann distribution function $\rho = \rho_{\rm c}\exp(2\varepsilon t-2\bar{\beta} M\Phi)$ into eq. (\ref{Poisson_2}), we get, 
\begin{align*}
    -\nabla^2\ln\left ( \frac{\rho}{\rho_{\rm c}} \right )^{1/2} = 4\pi GM^2\bar{\beta}\Psi^*\Psi - \bar{\beta} M\Lambda. 
\end{align*}
The left hand side in the above expression can be rewritten as, 
\begin{multline*}
    -\nabla^2\ln\left ( \frac{\rho}{\rho_{\rm c}} \right )^{1/2} = -\frac{\nabla^2|\Psi|}{|\Psi|} + \frac{\vec{\nabla}|\Psi|}{|\Psi|}\cdot\frac{\vec{\nabla}|\Psi|}{|\Psi|}\\
    = -\frac{\nabla^2|\Psi|}{|\Psi|} + M^2\vec{V}\cdot\vec{V},
\end{multline*}
where we have used $|\Psi|^2 = \rho$, $\vec{\nabla}\Phi = -\bar{\beta}^{-1}\vec{V}$. Hence, multiplying both (the right and the left hand) sides by $1/2M$ and equating them yields,
\begin{subequations}\label{Components_GP}
\begin{align}
    -\varepsilon = \frac{M}{2}\vec{V}\cdot\vec{V} - \frac{1}{2M}\frac{\nabla^2|\Psi|}{|\Psi|} - 2\pi GM\bar{\beta}\Psi^*\Psi,
\end{align}
where we have used $\varepsilon = \bar{\beta}\Lambda/2$ as in eq. (\ref{Fokker-Planck}). 
Moreover, Fick's \textit{second} law, $\partial_{\mu}(|\Psi|^2\xi^{\mu}) = 0$ where $\xi^{\mu} = (1, -\vec{V})$ yields the diffusion equation, 
\begin{align}
    \frac{\partial |\Psi|^2}{\partial t} = \vec{\nabla}\cdot(|\Psi|^2\vec{V}) = D\nabla^2|\Psi|^2.
\end{align}
\end{subequations}
Thus, we recognize eq. (\ref{Components_GP}) as simply the real and imaginary parts respectively of the Ginzburg-Landau (Gross-Pitaevskii) equation, 
\begin{align}\label{GL_eq}
  i\frac{\partial}{\partial t}\Psi = \left (-\frac{1}{2M}\nabla^2 + \frac{4\pi a_{\rm s}}{M}\Psi^*\Psi \right )\Psi,
\end{align}
where $\Psi = \sqrt{\rho}\exp(i(\varepsilon t - \bar{\beta}M\Phi))$ is the order parameter and $a_{\rm s} = -\bar{\beta} GM^2/2$ is the boson-boson scattering length. The minus sign signifies that the scattering is attractive in nature. Thus, $\Psi = \sqrt{\rho}\exp(iS)$ requires that,
\begin{align}
    S = \varepsilon t - \bar{\beta}M\Phi.
\end{align}
Consequently, the normalization condition, $\int_{\rm vol.} d^{\,3}x\,\Psi^*\Psi = n$ signifies that we are dealing with a theory of $n$ bosons. 

Finally, setting $i\partial\Psi/\partial t = E\Psi$ where $E$ is the chemical potential, the free energy, $\mathcal{F}$ which yields eq. (\ref{GL_eq}) under variation by $\Psi^*$ is given by, 
\begin{align}\label{free_energy}
    \mathcal{F} = \int d^{\,3}x \left ( \frac{-1}{2M}\vec{\nabla}\Psi^*\cdot\vec{\nabla}\Psi - U(|\Psi|^2) \right ),
\end{align}
where $U(|\Psi|^2) = -E|\Psi|^2 + 4\pi a_{\rm s}|\Psi|^4/2M$ is the well-known Mexican-hat potential which facilitates second-order phase transitions. Setting $\Psi\Psi^* = \rho_{\rm c}$ equal to the critical density, the first term in the free energy can be neglected. Thus, $\vec{\nabla}\Psi = 0$ and minimizing the potential $\partial U(\Psi\Psi^*)/\partial \Psi^* = 0$ with respect to $\Psi^*$ yields,
\begin{align}
    \Psi\left (E - \frac{4\pi a_{\rm s}}{M}\Psi\Psi^* \right ) = 0, 
\end{align}
where $E < 0$ and $a_{\rm s} < 0$. This minimization produces two solutions for the critical density, namely $\rho_{\rm c} = \Psi\Psi^* = 0$ and $\rho_{\rm c} = \Psi\Psi^* = ME/4\pi a_{\rm s} > 0$. Using $\bar{\beta} M = n$ given from eq. (\ref{equilibrium}), the scattering length becomes $a_{\rm s} = -MGn/2$, while the critical density with $-E \simeq \varepsilon = \bar{\beta} \Lambda/2$ yields,  
\begin{subequations}
\begin{align}\label{critical_2}
    \rho_{\rm c} = \frac{\bar{\beta} \Lambda M}{4\pi GMn} = \frac{\Lambda}{4\pi GM}.
\end{align}
Comparing eq. (\ref{density_profile}) with eq. (\ref{critical_2}), the constant $K$ is determined to be,
\begin{align}\label{K_eq}
    K = \bar{\beta} M \Lambda = n\Lambda.
\end{align}

\begin{figure}[!t]
\begin{center}
\includegraphics[width=\columnwidth,clip=true]{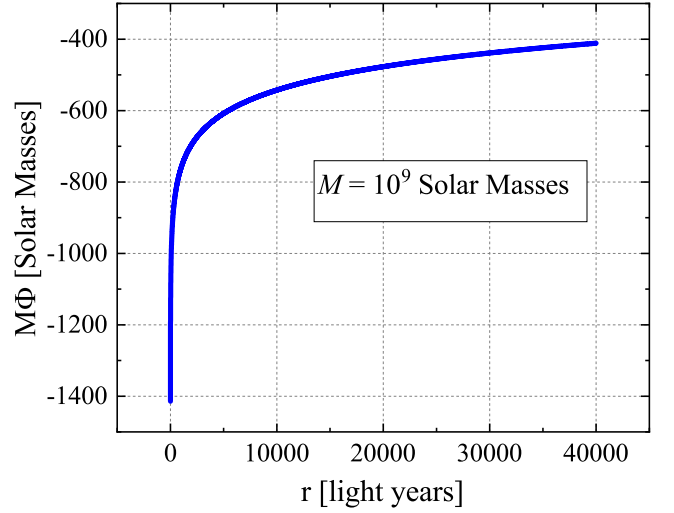}
 \caption{The logarithmic potential given in eq. (\ref{log_eq}) with $M = 10^9 \,M_\odot\ $
 }
\label{Potential_figure}
\end{center}
\end{figure}

The logarithmic potential $\Phi(r)$ in eq. (\ref{log_eq}) is binding (negative in value as shown in Fig. \ref{Potential_figure}) only when $r < 1/\sqrt{K} = 1/\sqrt{n\Lambda}$. This limits the sizes of galaxy halos to $r \leq 1/\sqrt{n\Lambda}$. Moreover, the critical density corresponds to the value of the number density at $r = r_{\rm c} = 1/\sqrt{n\Lambda} = v_{\rm c}/\sqrt{\Lambda}$ since,
\begin{align}
    \rho(r_{\rm c}) = \frac{1}{4\pi GMnr^2}\bigg\rvert_{r = r_{\rm c}} = \frac{\Lambda}{4\pi GM} = \rho_{\rm c}.
\end{align}
Thus, the maximum surface brightness $\Sigma_{\infty}$ of the galaxy allowed by the approach can be calculated using $M\rho_{\rm c}$ by integrating out the $z$ coordinate to yield, 
\begin{align}
    \Sigma_{\infty} = \int_{-z_{\infty} = -1/3a_{0}}^{+z_{\infty} = +1/3a_{0}} dz\,M\rho_{\rm c} = \frac{a_{0}}{2\pi G},
\end{align}
\end{subequations}
where the cut-off scale $z_{\infty} = 1/3a_{0}$ differs from the radius of the de-Sitter universe, $1/a_{0}$ by a factor of 3 arising from the number of dimensions.\footnote{Since $r = \sqrt{x^2 + y^2 + z^2}$, $r_{\infty} = 1/a_{0}$ implies that $z_{\infty} = 1/3a_{\infty}$, where the space is isotropic.} Consequently, we have arrived at the well-known Freeman law.\cite{freeman1970disks, mcgaugh1995galaxy} 

\red{

\subsection{Lagrangian Formalism}

For convenience, we perform the re-scaling, $|\Psi|^2$ $\rightarrow$ $|\Psi|^2/8\pi GM$, which ensures the complex-valued function has units, $[\Psi] = \rm energy$ or $\rm 1/length$, allowing the field equations to be compactly written as, $\nabla_{\mu}\mathcal{K}^{\mu}_{\,\,\nu} = \Psi^*\mathcal{D}_{\nu}\Psi$, where $\mathcal{K}_{\mu\nu} = R_{\mu\nu} + i2g\mathcal{F}_{\mu\nu}$. This means that the real part is simply the Bianchi identity, $\nabla^{\mu}R_{\mu\nu} = \frac{1}{2}\nabla_{\mu}R$, where $|\Psi|^2 = R$ and the imaginary part is obtained from varying the action with respect to the gauge field, $\mathcal{A}_{\mu}$ and subsequently replacing the partial derivative with a covariant one, $\partial_{\mu} \rightarrow \mathcal{D}_{\mu}$. Thus, we obtain an avenue to consider the Lagrangian density, $\mathcal{L}$ -- and hence the action $\mathcal{S} = \int d^{\,4}x\sqrt{-\det{(g_{\mu\nu})}}\,\mathcal{L}$ -- for the complex-valued function $\Psi$ in terms of a Lagrangian density of a complex scalar field coupled to the U$(1)$ gauge field $\mathcal{A}_{\mu}$ and a Riemann curvature term through the metric tensor,
\begin{multline}\label{Lagrangian_density_eq}
    \mathcal{L} =  - \frac{g^2}{2}\mathcal{F}_{\mu\nu}\mathcal{F}^{\mu\nu} - 
    \frac{1}{2}|\mathcal{D}_{\mu}\Psi|^2 + \frac{\bar{m}_{\rm P}^2}{2} |\Psi|^2 - \frac{1}{32}|\Psi|^4\\
    + P(\phi) -\frac{1}{32}R^{\mu\nu\sigma\rho}R_{\mu\nu\sigma\rho} + \frac{1}{8}\mathcal{K}^{\nu\mu}(\mathcal{K}_{\mu\nu})^\dagger,
\end{multline}
where $\bar{m}_{\rm P} = 1/\sqrt{8\pi G}$ is the reduced Planck mass, $\mathcal{K}_{\mu\nu} = (\mathcal{K}_{\nu\mu})^* = (\mathcal{K}_{\mu\nu})^{\dagger}$ is Hermitian and,
\begin{subequations}
\begin{align}
    P(\phi) = \frac{1}{2}(\partial_{\mu}\phi)^2 - V(\phi),\\
    M\rho(\phi) = \frac{1}{2}(\partial_{\mu}\phi)^2 + V(\phi)
\end{align}
\end{subequations}
where $\phi$ is a real scalar field and $V(\phi)$ is its potential, both generically employed here solely to generate the uncharged pressure-less source in eq. (\ref{eq_of_state})), with the constraints, $P(\phi) = 0$ and $u_{\mu} = \partial_{\mu}\phi/\sqrt{-(\partial_{\nu}\phi)^2}$. \footnote{\red{To simplify the expressions, one can re-scale $\phi \rightarrow \bar{m}_{\rm P}\phi$, and then make the choice, $\phi = M\int u_{\mu}dx^{\mu}$, which yields the geodesic equation when varied with respect to $x^{\mu}$.}} The cosmological constant, $\Lambda$ can be generated in a similar manner with the constraint $P(\phi) = -M\rho(\phi)$ and $V(\phi) = \bar{m}_{\rm P}^2\Lambda$.

At first glance, the action appears to have a completely different form from the Einstein-Hilbert action we expect from Einstein's General Relativity. Nonetheless, we can recognize the Ginzburg-Landau Lagrangian in the first line. Proceeding, we plug in $\mathcal{K}_{\mu\nu} = R_{\mu\nu} + i2g\mathcal{F}_{\mu\nu}$ and $\Psi = \sqrt{R}\exp(iS) \propto \exp(-n\Phi)\exp(iS)$, and then use $\mathcal{F}_{\mu\nu}R^{\mu\nu} = 0$, $\mathcal{D}_{\mu} = \partial_{\mu} - ig\mathcal{A}_{\mu}$, $\partial_{\mu}S = M\xi_{\mu} + g\mathcal{A}_{\mu}$, $\partial_{\mu}\Phi = \bar{\beta}^{-1}\xi_{\mu}$, $n = \bar{\beta}M$ and $\xi^{\mu}\xi_{\mu} = 0$ to find,
\begin{align}
    \mathcal{L} = \mathcal{L}_{\rm EH} + P + \mathcal{L}_{\rm GB},
\end{align}
where $\mathcal{L}_{\rm EH} = \bar{m}_{\rm P}^2R/2$ is the Einstein-Hilbert Lagrangian density and $\mathcal{L}_{\rm GB} = -\frac{1}{32}(R^{\mu\nu\sigma\rho}R_{\mu\nu\sigma\rho} - 4R^{\mu\nu}R_{\mu\nu} + R^2)$ is proportional to the Pfaffian of the Riemann curvature two-form, which is topological in nature, \textit{i.e.} it does not contribute to the Einstein Field Equations since it is a total derivative which yields the Euler characteristic, $\chi (h) = -(4/\pi^2)\int_{\mathcal{M}} d^{\,4}x\sqrt{-\det{(g_{\mu\nu})}}\,\mathcal{L}_{\rm GB}$ of the manifold, $\mathcal{M}$.\cite{buzano2019higher, lovelock1971einstein} In fact, this observation, together with the constraint provided by the Hermitian tensor equation, fix \textit{a priori} all the coefficients in eq. (\ref{Lagrangian_density_eq}).

Thus, due to the presence of the Hermitian tensor in the Lagrangian density, the stress-energy-momentum tensor of the novel gauge field $\mathcal{A}_{\mu}$ \textit{need not couple to the Einstein Field Equations nor to dark matter}, as expected from eq. (\ref{geodesic_1}). Finally, the fact that $\Psi$ and $\mathcal{K}_{\mu\nu}$ terms in the Lagrangian density transform into Gauss-Bonnet gravity terms under the constraint $\nabla_{\mu}K^{\mu}_{\,\,\nu} = \Psi^*\mathcal{D}_{\nu}\Psi$ attests to its utility in capturing the topological aspects of gravity. This will be explored in future publications.\cite{Kanyolo2021}

}

\subsection{Metric Solution}

We offer a suitable relativistic solution of eq. (\ref{eq_of_state}) with the constraint $\rho = \rho_{\rm c}\exp(-2n\Phi)$, which we show reproduces the result in eq. (\ref{log_eq}) in the Newtonian limit. A metric solution is useful in determining null geodesics used to establish the existence of dark matter through gravitational lensing. We shall consider spherically symmetric matter whose metric solution is given by the generic form,\cite{mak2013isotropic, thorne2000gravitation}
\begin{align}\label{metric}
    d\tau^2 = \exp(2\Phi(r))dt^2 - \frac{dr^2}{1 - 2Gm(r)/r} - r^2d\Omega^2
\end{align}
where $d\Omega^2 = d\theta^2 + \sin^2\theta d\varphi^2$ is the metric of the 2-sphere. This generic metric takes the same form as the interior solution of spherically symmetric dust, where $g_{00} = \xi^{\mu}\xi_{\mu} = u^{\mu}u_{\mu}\exp(2\Phi) = -\exp(2\Phi)$ is fixed by $\xi^{\mu} = (1, \vec{0})$ and $u^{\mu}u_{\mu} = -1$. 

Due to the symmetry of the problem, we need only consider the $t-t$ and the $r-r$ components of eq. (\ref{eq_of_state}), together with the conservation of energy-momentum, $\nabla_{\mu}(\rho u^{\mu}u^{\nu}) = 0$. The latter merely yields the geodesic equation, $u^{\mu}\nabla_{\mu}u^{\nu} = 0$ and the continuity equation $\nabla_{\mu}(\rho u^{\mu}) = 0$. On the other hand, plugging in eq. (\ref{metric}) and $u^{\mu} = \exp(-\Phi)\xi^{\mu}$, where $\xi^{\mu} = (1, \vec{0})$ is time-like, the components respectively become, \begin{subequations}\label{metric_solutions}
\begin{align}\label{time_time}
    \frac{2G}{r^2}\frac{\partial m(r)}{\partial r} = 8\pi G M\rho(r) + \Lambda,
\end{align}
\begin{align*}
    \frac{2}{r}\left ( 1 - \frac{2Gm(r)}{r} \right )\frac{\partial \Phi}{\partial r} - \frac{2Gm(r)}{r^3} = -\Lambda,
\end{align*}
where we recall that $\rho(r) = \rho_{\rm c}\exp(-2n\Phi(r))$. Thus, the functions $\Phi(r)$ and $m(r)$ are coupled to each other, namely when the mass $m(r)$ varies in the first equation, it determines the function $\Phi(r)$ through the number density $\rho(r)$, whereas when the function $\Phi(r)$ varies in the second equation, it determines the mass, $m(r)$. Differentiating by $r$, the second equation yields,  
\begin{multline}\label{space_space}
    \frac{1}{r^2}\frac{\partial}{\partial r}\left ( r^2\left [ 1 - \frac{2Gm(r)}{r} \right ]\frac{\partial \Phi(r)}{\partial r} \right )\\
    = \frac{G}{r^2}\frac{\partial m(r)}{\partial r} - \frac{3}{2}\Lambda = 4\pi GM\rho_{\rm c}\exp(-2n\Phi(r)) - \Lambda,
\end{multline}
\end{subequations}
where we have used eq. (\ref{time_time}) in the last line. These equations (eq. (\ref{metric_solutions})) basically correspond to the relativistic versions of eq. (\ref{Poisson_2}) and eq. (\ref{dark_matter}) respectively. 

In fact, plugging in the trial function,
\begin{align}\label{trial_function_eq}
    \Phi(r) = \frac{1}{2}\ln (1 - 2Gm(r)/r),
\end{align}
into eq. (\ref{space_space}) yields the following mass equations, 
\begin{subequations}\label{mass_eq}
\begin{align}\label{mass_eq1}
    \frac{G}{r^2}\frac{\partial m(r)}{\partial r} - \frac{\Lambda}{2}
    = 4\pi G M\rho_{\rm c}\exp\left (-n\ln \left (1-\frac{2Gm(r)}{r} \right )\right ),
\end{align}
and,
\begin{align}\label{mass_eq2}
    \frac{G}{r^2}\frac{\partial}{\partial r} \left (r\frac{\partial m(r)}{\partial r} \right ) - \frac{3}{2}\Lambda = 0,
\end{align}    
\end{subequations}
useful in further simplifying eq. (\ref{metric_solutions}). The general solution for eq. (\ref{mass_eq2}) is given by,
\begin{subequations}
\begin{align}\label{solution_eq}
    m(r) = M + 3\mathcal{M}\ln (\alpha r) + \frac{\Lambda r^3}{6G},
\end{align}
where $\mathcal{M}$ and $\alpha$ are undetermined constants. \red{Plugging this solution into the left-hand side of eq. (\ref{mass_eq1}), requires that $M\rho (r) = 3\mathcal{M}/4\pi r^3 = M\rho_{\rm c}\exp(-2n\ln(4\pi r^3M\rho_{\rm c}/3\mathcal{M})^{1/2n})$. However, this appears to contradict the constraint on the right-hand side ($\rho(r) \neq \rho_{\rm c}\exp(-2n\Phi(r))$)} unless $\mathcal{M} = 0$, where the solution given by eq. (\ref{solution_eq}) corresponds to the de Sitter-Schwarzschild metric.\cite{nariai1950some, nariai1999new} \red{Nonetheless, we recall the constraint was brought about by our novel approach, \textit{i.e.} this constraint need not be considered in Einstein's General Relativity. Thus, such a profile would correspond to the asymptotic limit of the Navarro-Frenk-White (NFW) profile, where $\mathcal{M} = 4\pi r_{\rm s}^3M\rho_{\rm 0}/3$, and the scale radius $r_{\rm s} \ll r$ and the central density, $M\rho_{0}$ vary from halo to halo.} 
\red{On the other hand, imposing this constraint,} we can make progress by performing a suitable approximation to arrive at a useful expression. Since in the limit $2Gm(r)/r \ll 1$, eq. (\ref{space_space}) reduces to the Emden-Chandrasekhar equation (given by eq. (\ref{Emden-Chandrasekhar_eq})), the solution and its Taylor-Maclaurin approximation respectively can be taken to be, 
\begin{multline}\label{approx_solution_eq}
    m(r) = r/2Gn - (r/2Gn)\ln(-Kr^2/3)\\
    \simeq r/nG + \Lambda r^3/6G,
\end{multline}
\end{subequations}
where $K = n\Lambda$ and the critical density $\rho_{\rm c}$ has been renormalized to $\rho_{\rm c}^* = -\rho_{\rm c}\exp(1 + \ln (3))$\red{, using $\exp(i\pi) = -1$ with $\rho_{\rm c} < 0$ in order for $\rho_{\rm c}^* > 0$. Thus, the approximation also corresponds to the Taylor-Maclaurin approximation of eq. (\ref{solution_eq}) when $\alpha = 1/3\mathcal{M}Gn$ and $\rho_0 = 1/4\pi r_{\rm s}^3$}.

\section{Discussion}

Based on the discussion in the phase transition section, the Newtonian potential $\Phi$ should take on two forms when $\Psi\Psi^* = 0$ and $\Psi\Psi^* \neq 0$, which corresponds to the Newtonian solution with $\Lambda = 0$ and the Modified Newtonian solution with $\Lambda \neq 0$ respectively according to eq. (\ref{critical_2}). This is equivalent to the limits $1/n \rightarrow 0$ and $1/n \neq 0$ respectively, where $n = 1/\sqrt{GMa_0}$ and $a_{0} = \sqrt{\Lambda/3}$. In essence, this corresponds to considering whether the size of the universe is relevant to the problem at hand.

In particular, the Tully-Fisher relation is reproduced in a straightforward manner for $1/n \neq 0$ following from the normalization condition $\int_{\rm vol.} d^{\,3}x \, \Psi^*\Psi = n$, where $\Psi = \sqrt{\rho}\exp(iS)$ is the order parameter. On the other hand, in the limit $1/n \rightarrow 0$, the critical density approaches zero ($\rho_{\rm c} \rightarrow 0$) and eq. (\ref{Poisson_2}) yields the inverse-square law, $\Phi(r) = GM/r$ as expected. Thus, for a small amount of baryonic mass ($M \ll 1/Ga_0$) such as the solar system and in some ultra-diffuse galaxies\cite{van2018galaxy, van2016high}, taking $1/n \rightarrow 0$ is appropriate since it corresponds to taking the size of the universe to be considerably larger than the system. This renders the effects of dark matter (and hence the Tully-Fisher relation) absent in such systems. However, as the amount of baryonic mass becomes large ($M \rightarrow 1/Ga_0$), the inverse square-law is no longer appropriate since the size of the system becomes comparable to the finite size of the de Sitter universe. This means that the critical density, $\rho_{\rm c}$ can no longer be neglected. This corresponds to the dynamics described by the Emden-Chandrasekhar equation given in eq. (\ref{Emden-Chandrasekhar_eq}), equivalent to a phase transition $\Psi^*\Psi \neq 0$, which corresponds to a deviation from the inverse square law, where the dark matter within radius $r$ is given by $m(r) = M\int_{0}^{r} dr'\,4\pi r'^2 \,\rho(r')$ under the constraint, $\rho(r) =\rho_{\rm c}\exp(-2n\Phi(r)) = \Psi^*\Psi$. \red{This constraint is not necessarily available in Einstein's General Relativity. Relaxing this constraint yields a density profile compatible with the asymptotic behavior of the NFW profile.}

Note that eq. (\ref{GL_eq}) differs from the conventional superfluid dark matter approaches which are essentially Ginzburg-Landau (Gross-Pitaevskii) theories where the $|\Psi|^4$ term (two particle interaction term) of the potential $V(|\Psi|^2)$ in the free energy function in eq. (\ref{free_energy}) is replaced by a three-particle interaction term.\cite{berezhiani2015theory, sharma2019equation} Such approaches necessarily differ from our approach herein since they are not only non-relativistic, but do not readily predict the connection of $a_{0}$ to the cosmological constant. A precise analysis using instead the relativistic solution of eq. (\ref{eq_of_state}), re-affirms the aforementioned conclusions. 

Finally, that $2n$ can be interpreted as the number of independent affine transformations given by eq. (\ref{n_times}) implies that a finite universe renders this number finite. This point may prove to be more profound than we have explored herein.

\section{Conclusion}

We \red{have proposed} a novel approach to the relativistic equations of gravity \red{which introduces a novel constraint} in Einstein’s general relativity, where the Einstein Field Equations and the geodesic equations are effectively interpreted as relativistic Fokker-Planck and Langevin equations respectively, satisfying the well-known fluctuation-dissipation theorem (Einstein-Smoluchowski equation). \red{This naturally reproduces an isothermal dark matter profile at the asymptotic regions of galaxies consistent with the Tully-Fisher relation}. The proposed equations not only contain a complex-valued function whose role is shown to be the order parameter for the dark matter within the context of Ginzburg-Landau theory, but also involve all the fundamental constants of nature when written in System International (SI) units, namely Planck's constant $\hbar$, Boltzmann constant $k_{\rm B}$, gravitational constant $G$ and speed of light in vacuum $c$. Thus, we present a theoretical approach that will indubitably have significant implications on the \red{cold dark matter paradigm}
since we provide a framework where the Tully-Fisher relation in galaxies is satisfied within the context of general relativity.

The authors would like to acknowledge the financial support of TEPCO Memorial Foundation, Japan Society for the Promotion of Science (JSPS KAKENHI Grant Numbers 19K15685 and 21K14730) and Japan Prize Foundation. The authors also acknowledge fruitful discussions with D. Ntara during the cradle of the ideas herein, and the proofreading work on the manuscript done by Edfluent. Both authors are grateful for the unwavering support from their family members (T. M.: Ishii Family, Sakaguchi Family and Masese Family; G. M. K.: Ngumbi Family).  
 
\bibliography{Novel_Constraint}

\providecommand{\noopsort}[1]{}\providecommand{\singleletter}[1]{#1}%
\begin{thebibliography}{95}%
\makeatletter
\providecommand \@ifxundefined [1]{%
 \@ifx{#1\undefined}
}%
\providecommand \@ifnum [1]{%
 \ifnum #1\expandafter \@firstoftwo
 \else \expandafter \@secondoftwo
 \fi
}%
\providecommand \@ifx [1]{%
 \ifx #1\expandafter \@firstoftwo
 \else \expandafter \@secondoftwo
 \fi
}%
\providecommand \natexlab [1]{#1}%
\providecommand \enquote  [1]{``#1''}%
\providecommand \bibnamefont  [1]{#1}%
\providecommand \bibfnamefont [1]{#1}%
\providecommand \citenamefont [1]{#1}%
\providecommand \href@noop [0]{\@secondoftwo}%
\providecommand \href [0]{\begingroup \@sanitize@url \@href}%
\providecommand \@href[1]{\@@startlink{#1}\@@href}%
\providecommand \@@href[1]{\endgroup#1\@@endlink}%
\providecommand \@sanitize@url [0]{\catcode `\\12\catcode `\$12\catcode
  `\&12\catcode `\#12\catcode `\^12\catcode `\_12\catcode `\%12\relax}%
\providecommand \@@startlink[1]{}%
\providecommand \@@endlink[0]{}%
\providecommand \url  [0]{\begingroup\@sanitize@url \@url }%
\providecommand \@url [1]{\endgroup\@href {#1}{\urlprefix }}%
\providecommand \urlprefix  [0]{URL }%
\providecommand \Eprint [0]{\href }%
\providecommand \doibase [0]{https://doi.org/}%
\providecommand \selectlanguage [0]{\@gobble}%
\providecommand \bibinfo  [0]{\@secondoftwo}%
\providecommand \bibfield  [0]{\@secondoftwo}%
\providecommand \translation [1]{[#1]}%
\providecommand \BibitemOpen [0]{}%
\providecommand \bibitemStop [0]{}%
\providecommand \bibitemNoStop [0]{.\EOS\space}%
\providecommand \EOS [0]{\spacefactor3000\relax}%
\providecommand \BibitemShut  [1]{\csname bibitem#1\endcsname}%
\let\auto@bib@innerbib\@empty
\bibitem [{\citenamefont {Famaey}\ and\ \citenamefont
  {McGaugh}(2012)}]{famaey2012modified}%
  \BibitemOpen
  \bibfield  {author} {\bibinfo {author} {\bibfnamefont {B.}~\bibnamefont
  {Famaey}}\ and\ \bibinfo {author} {\bibfnamefont {S.~S.}\ \bibnamefont
  {McGaugh}},\ }\bibfield  {title} {\bibinfo {title} {Modified newtonian
  dynamics (mond): observational phenomenology and relativistic extensions},\
  }\href@noop {} {\bibfield  {journal} {\bibinfo  {journal} {Living Reviews in
  Relativity}\ }\textbf {\bibinfo {volume} {15}},\ \bibinfo {pages} {10}
  (\bibinfo {year} {2012})}\BibitemShut {NoStop}%
\bibitem [{\citenamefont {Zwicky}(1937)}]{zwicky1937masses}%
  \BibitemOpen
  \bibfield  {author} {\bibinfo {author} {\bibfnamefont {F.}~\bibnamefont
  {Zwicky}},\ }\bibfield  {title} {\bibinfo {title} {On the masses of nebulae
  and of clusters of nebulae},\ }\href@noop {} {\bibfield  {journal} {\bibinfo
  {journal} {The Astrophysical Journal}\ }\textbf {\bibinfo {volume} {86}},\
  \bibinfo {pages} {217} (\bibinfo {year} {1937})}\BibitemShut {NoStop}%
\bibitem [{\citenamefont {Babcock}(1939)}]{babcock1939rotation}%
  \BibitemOpen
  \bibfield  {author} {\bibinfo {author} {\bibfnamefont {H.~W.}\ \bibnamefont
  {Babcock}},\ }\bibfield  {title} {\bibinfo {title} {The rotation of the
  andromeda nebula},\ }\href@noop {} {\bibfield  {journal} {\bibinfo  {journal}
  {Lick Observatory Bulletin}\ }\textbf {\bibinfo {volume} {19}},\ \bibinfo
  {pages} {41} (\bibinfo {year} {1939})}\BibitemShut {NoStop}%
\bibitem [{\citenamefont {Freeman}(1970)}]{freeman1970disks}%
  \BibitemOpen
  \bibfield  {author} {\bibinfo {author} {\bibfnamefont {K.~C.}\ \bibnamefont
  {Freeman}},\ }\bibfield  {title} {\bibinfo {title} {On the disks of spiral
  and s0 galaxies},\ }\href@noop {} {\bibfield  {journal} {\bibinfo  {journal}
  {The Astrophysical Journal}\ }\textbf {\bibinfo {volume} {160}},\ \bibinfo
  {pages} {811} (\bibinfo {year} {1970})}\BibitemShut {NoStop}%
\bibitem [{\citenamefont {Rubin}\ and\ \citenamefont
  {Ford~Jr}(1970)}]{rubin1970rotation}%
  \BibitemOpen
  \bibfield  {author} {\bibinfo {author} {\bibfnamefont {V.~C.}\ \bibnamefont
  {Rubin}}\ and\ \bibinfo {author} {\bibfnamefont {W.~K.}\ \bibnamefont
  {Ford~Jr}},\ }\bibfield  {title} {\bibinfo {title} {Rotation of the andromeda
  nebula from a spectroscopic survey of emission regions},\ }\href@noop {}
  {\bibfield  {journal} {\bibinfo  {journal} {The Astrophysical Journal}\
  }\textbf {\bibinfo {volume} {159}},\ \bibinfo {pages} {379} (\bibinfo {year}
  {1970})}\BibitemShut {NoStop}%
\bibitem [{\citenamefont {Rogstad}\ and\ \citenamefont
  {Shostak}(1972)}]{rogstad1972gross}%
  \BibitemOpen
  \bibfield  {author} {\bibinfo {author} {\bibfnamefont {D.}~\bibnamefont
  {Rogstad}}\ and\ \bibinfo {author} {\bibfnamefont {G.}~\bibnamefont
  {Shostak}},\ }\bibfield  {title} {\bibinfo {title} {Gross properties of five
  scd galaxies as determined from 21-centimeter observations},\ }\href@noop {}
  {\bibfield  {journal} {\bibinfo  {journal} {The Astrophysical Journal}\
  }\textbf {\bibinfo {volume} {176}},\ \bibinfo {pages} {315} (\bibinfo {year}
  {1972})}\BibitemShut {NoStop}%
\bibitem [{\citenamefont {Corbelli}\ and\ \citenamefont
  {Salucci}(2000)}]{corbelli2000extended}%
  \BibitemOpen
  \bibfield  {author} {\bibinfo {author} {\bibfnamefont {E.}~\bibnamefont
  {Corbelli}}\ and\ \bibinfo {author} {\bibfnamefont {P.}~\bibnamefont
  {Salucci}},\ }\bibfield  {title} {\bibinfo {title} {The extended rotation
  curve and the dark matter halo of m33},\ }\href@noop {} {\bibfield  {journal}
  {\bibinfo  {journal} {Monthly Notices of the Royal Astronomical Society}\
  }\textbf {\bibinfo {volume} {311}},\ \bibinfo {pages} {441} (\bibinfo {year}
  {2000})}\BibitemShut {NoStop}%
\bibitem [{\citenamefont {Faber}\ and\ \citenamefont
  {Jackson}(1976)}]{faber1976velocity}%
  \BibitemOpen
  \bibfield  {author} {\bibinfo {author} {\bibfnamefont {S.}~\bibnamefont
  {Faber}}\ and\ \bibinfo {author} {\bibfnamefont {R.~E.}\ \bibnamefont
  {Jackson}},\ }\bibfield  {title} {\bibinfo {title} {Velocity dispersions and
  mass-to-light ratios for elliptical galaxies},\ }\href@noop {} {\bibfield
  {journal} {\bibinfo  {journal} {The Astrophysical Journal}\ }\textbf
  {\bibinfo {volume} {204}},\ \bibinfo {pages} {668} (\bibinfo {year}
  {1976})}\BibitemShut {NoStop}%
\bibitem [{\citenamefont {Trimble}(1987)}]{trimble1987existence}%
  \BibitemOpen
  \bibfield  {author} {\bibinfo {author} {\bibfnamefont {V.}~\bibnamefont
  {Trimble}},\ }\bibfield  {title} {\bibinfo {title} {Existence and nature of
  dark matter in the universe},\ }\href@noop {} {\bibfield  {journal} {\bibinfo
   {journal} {Annual Review of Astronomy and Astrophysics}\ }\textbf {\bibinfo
  {volume} {25}},\ \bibinfo {pages} {425} (\bibinfo {year} {1987})}\BibitemShut
  {NoStop}%
\bibitem [{\citenamefont {Bertone}\ \emph {et~al.}(2005)\citenamefont
  {Bertone}, \citenamefont {Hooper},\ and\ \citenamefont
  {Silk}}]{bertone2005particle}%
  \BibitemOpen
  \bibfield  {author} {\bibinfo {author} {\bibfnamefont {G.}~\bibnamefont
  {Bertone}}, \bibinfo {author} {\bibfnamefont {D.}~\bibnamefont {Hooper}},\
  and\ \bibinfo {author} {\bibfnamefont {J.}~\bibnamefont {Silk}},\ }\bibfield
  {title} {\bibinfo {title} {Particle dark matter: evidence, candidates and
  constraints},\ }\href@noop {} {\bibfield  {journal} {\bibinfo  {journal}
  {Physics Reports}\ }\textbf {\bibinfo {volume} {405}},\ \bibinfo {pages}
  {279} (\bibinfo {year} {2005})}\BibitemShut {NoStop}%
\bibitem [{\citenamefont {Copi}\ \emph {et~al.}(1995)\citenamefont {Copi},
  \citenamefont {Schramm},\ and\ \citenamefont {Turner}}]{copi1995big}%
  \BibitemOpen
  \bibfield  {author} {\bibinfo {author} {\bibfnamefont {C.~J.}\ \bibnamefont
  {Copi}}, \bibinfo {author} {\bibfnamefont {D.~N.}\ \bibnamefont {Schramm}},\
  and\ \bibinfo {author} {\bibfnamefont {M.~S.}\ \bibnamefont {Turner}},\
  }\bibfield  {title} {\bibinfo {title} {Big-bang nucleosynthesis and the
  baryon density of the universe},\ }\href@noop {} {\bibfield  {journal}
  {\bibinfo  {journal} {Science}\ }\textbf {\bibinfo {volume} {267}},\ \bibinfo
  {pages} {192} (\bibinfo {year} {1995})}\BibitemShut {NoStop}%
\bibitem [{\citenamefont {Clowe}\ \emph {et~al.}(2006)\citenamefont {Clowe},
  \citenamefont {Brada{\v{c}}}, \citenamefont {Gonzalez}, \citenamefont
  {Markevitch}, \citenamefont {Randall}, \citenamefont {Jones},\ and\
  \citenamefont {Zaritsky}}]{clowe2006direct}%
  \BibitemOpen
  \bibfield  {author} {\bibinfo {author} {\bibfnamefont {D.}~\bibnamefont
  {Clowe}}, \bibinfo {author} {\bibfnamefont {M.}~\bibnamefont {Brada{\v{c}}}},
  \bibinfo {author} {\bibfnamefont {A.~H.}\ \bibnamefont {Gonzalez}}, \bibinfo
  {author} {\bibfnamefont {M.}~\bibnamefont {Markevitch}}, \bibinfo {author}
  {\bibfnamefont {S.~W.}\ \bibnamefont {Randall}}, \bibinfo {author}
  {\bibfnamefont {C.}~\bibnamefont {Jones}},\ and\ \bibinfo {author}
  {\bibfnamefont {D.}~\bibnamefont {Zaritsky}},\ }\bibfield  {title} {\bibinfo
  {title} {A direct empirical proof of the existence of dark matter},\
  }\href@noop {} {\bibfield  {journal} {\bibinfo  {journal} {The Astrophysical
  Journal Letters}\ }\textbf {\bibinfo {volume} {648}},\ \bibinfo {pages}
  {L109} (\bibinfo {year} {2006})}\BibitemShut {NoStop}%
\bibitem [{\citenamefont {Aguirre}\ \emph {et~al.}(2001)\citenamefont
  {Aguirre}, \citenamefont {Schaye},\ and\ \citenamefont
  {Quataert}}]{aguirre2001problems}%
  \BibitemOpen
  \bibfield  {author} {\bibinfo {author} {\bibfnamefont {A.}~\bibnamefont
  {Aguirre}}, \bibinfo {author} {\bibfnamefont {J.}~\bibnamefont {Schaye}},\
  and\ \bibinfo {author} {\bibfnamefont {E.}~\bibnamefont {Quataert}},\
  }\bibfield  {title} {\bibinfo {title} {Problems for modified newtonian
  dynamics in clusters and the ly$\alpha$ forest?},\ }\href@noop {} {\bibfield
  {journal} {\bibinfo  {journal} {The Astrophysical Journal}\ }\textbf
  {\bibinfo {volume} {561}},\ \bibinfo {pages} {550} (\bibinfo {year}
  {2001})}\BibitemShut {NoStop}%
\bibitem [{\citenamefont {Van~Dokkum}\ \emph {et~al.}(2018)\citenamefont
  {Van~Dokkum}, \citenamefont {Danieli}, \citenamefont {Cohen}, \citenamefont
  {Merritt}, \citenamefont {Romanowsky}, \citenamefont {Abraham}, \citenamefont
  {Brodie}, \citenamefont {Conroy}, \citenamefont {Lokhorst}, \citenamefont
  {Mowla} \emph {et~al.}}]{van2018galaxy}%
  \BibitemOpen
  \bibfield  {author} {\bibinfo {author} {\bibfnamefont {P.}~\bibnamefont
  {Van~Dokkum}}, \bibinfo {author} {\bibfnamefont {S.}~\bibnamefont {Danieli}},
  \bibinfo {author} {\bibfnamefont {Y.}~\bibnamefont {Cohen}}, \bibinfo
  {author} {\bibfnamefont {A.}~\bibnamefont {Merritt}}, \bibinfo {author}
  {\bibfnamefont {A.~J.}\ \bibnamefont {Romanowsky}}, \bibinfo {author}
  {\bibfnamefont {R.}~\bibnamefont {Abraham}}, \bibinfo {author} {\bibfnamefont
  {J.}~\bibnamefont {Brodie}}, \bibinfo {author} {\bibfnamefont
  {C.}~\bibnamefont {Conroy}}, \bibinfo {author} {\bibfnamefont
  {D.}~\bibnamefont {Lokhorst}}, \bibinfo {author} {\bibfnamefont
  {L.}~\bibnamefont {Mowla}}, \emph {et~al.},\ }\bibfield  {title} {\bibinfo
  {title} {A galaxy lacking dark matter},\ }\href@noop {} {\bibfield  {journal}
  {\bibinfo  {journal} {Nature}\ }\textbf {\bibinfo {volume} {555}},\ \bibinfo
  {pages} {629} (\bibinfo {year} {2018})}\BibitemShut {NoStop}%
\bibitem [{\citenamefont {Van~Dokkum}\ \emph {et~al.}(2016)\citenamefont
  {Van~Dokkum}, \citenamefont {Abraham}, \citenamefont {Brodie}, \citenamefont
  {Conroy}, \citenamefont {Danieli}, \citenamefont {Merritt}, \citenamefont
  {Mowla}, \citenamefont {Romanowsky},\ and\ \citenamefont
  {Zhang}}]{van2016high}%
  \BibitemOpen
  \bibfield  {author} {\bibinfo {author} {\bibfnamefont {P.}~\bibnamefont
  {Van~Dokkum}}, \bibinfo {author} {\bibfnamefont {R.}~\bibnamefont {Abraham}},
  \bibinfo {author} {\bibfnamefont {J.}~\bibnamefont {Brodie}}, \bibinfo
  {author} {\bibfnamefont {C.}~\bibnamefont {Conroy}}, \bibinfo {author}
  {\bibfnamefont {S.}~\bibnamefont {Danieli}}, \bibinfo {author} {\bibfnamefont
  {A.}~\bibnamefont {Merritt}}, \bibinfo {author} {\bibfnamefont
  {L.}~\bibnamefont {Mowla}}, \bibinfo {author} {\bibfnamefont
  {A.}~\bibnamefont {Romanowsky}},\ and\ \bibinfo {author} {\bibfnamefont
  {J.}~\bibnamefont {Zhang}},\ }\bibfield  {title} {\bibinfo {title} {A high
  stellar velocity dispersion and~ 100 globular clusters for the ultra-diffuse
  galaxy dragonfly 44},\ }\href@noop {} {\bibfield  {journal} {\bibinfo
  {journal} {The Astrophysical Journal Letters}\ }\textbf {\bibinfo {volume}
  {828}},\ \bibinfo {pages} {L6} (\bibinfo {year} {2016})}\BibitemShut
  {NoStop}%
\bibitem [{\citenamefont {Carroll}\ and\ \citenamefont
  {Ostlie}(2017)}]{carroll2017introduction}%
  \BibitemOpen
  \bibfield  {author} {\bibinfo {author} {\bibfnamefont {B.~W.}\ \bibnamefont
  {Carroll}}\ and\ \bibinfo {author} {\bibfnamefont {D.~A.}\ \bibnamefont
  {Ostlie}},\ }\href@noop {} {\emph {\bibinfo {title} {An introduction to
  modern astrophysics}}}\ (\bibinfo  {publisher} {Cambridge University Press},\
  \bibinfo {year} {2017})\BibitemShut {NoStop}%
\bibitem [{\citenamefont {Huo}\ \emph {et~al.}(2020)\citenamefont {Huo},
  \citenamefont {Yu},\ and\ \citenamefont {Zhong}}]{huo2020structure}%
  \BibitemOpen
  \bibfield  {author} {\bibinfo {author} {\bibfnamefont {R.}~\bibnamefont
  {Huo}}, \bibinfo {author} {\bibfnamefont {H.-B.}\ \bibnamefont {Yu}},\ and\
  \bibinfo {author} {\bibfnamefont {Y.-M.}\ \bibnamefont {Zhong}},\ }\bibfield
  {title} {\bibinfo {title} {The structure of dissipative dark matter halos},\
  }\href@noop {} {\bibfield  {journal} {\bibinfo  {journal} {Journal of
  Cosmology and Astroparticle Physics}\ }\textbf {\bibinfo {volume}
  {2020}}\bibinfo  {number} { (06)},\ \bibinfo {pages} {051}}\BibitemShut
  {NoStop}%
\bibitem [{\citenamefont {Essig}\ \emph {et~al.}(2019)\citenamefont {Essig},
  \citenamefont {McDermott}, \citenamefont {Yu},\ and\ \citenamefont
  {Zhong}}]{essig2019constraining}%
  \BibitemOpen
\bibfield  {number} {  }\bibfield  {author} {\bibinfo {author} {\bibfnamefont
  {R.}~\bibnamefont {Essig}}, \bibinfo {author} {\bibfnamefont {S.~D.}\
  \bibnamefont {McDermott}}, \bibinfo {author} {\bibfnamefont {H.-B.}\
  \bibnamefont {Yu}},\ and\ \bibinfo {author} {\bibfnamefont {Y.-M.}\
  \bibnamefont {Zhong}},\ }\bibfield  {title} {\bibinfo {title} {Constraining
  dissipative dark matter self-interactions},\ }\href@noop {} {\bibfield
  {journal} {\bibinfo  {journal} {Physical Review Letters}\ }\textbf {\bibinfo
  {volume} {123}},\ \bibinfo {pages} {121102} (\bibinfo {year}
  {2019})}\BibitemShut {NoStop}%
\bibitem [{\citenamefont {McGaugh}\ \emph {et~al.}(2000)\citenamefont
  {McGaugh}, \citenamefont {Schombert}, \citenamefont {Bothun},\ and\
  \citenamefont {De~Blok}}]{mcgaugh2000baryonic}%
  \BibitemOpen
  \bibfield  {author} {\bibinfo {author} {\bibfnamefont {S.~S.}\ \bibnamefont
  {McGaugh}}, \bibinfo {author} {\bibfnamefont {J.~M.}\ \bibnamefont
  {Schombert}}, \bibinfo {author} {\bibfnamefont {G.~D.}\ \bibnamefont
  {Bothun}},\ and\ \bibinfo {author} {\bibfnamefont {W.}~\bibnamefont
  {De~Blok}},\ }\bibfield  {title} {\bibinfo {title} {The baryonic tully-fisher
  relation},\ }\href@noop {} {\bibfield  {journal} {\bibinfo  {journal} {The
  Astrophysical Journal Letters}\ }\textbf {\bibinfo {volume} {533}},\ \bibinfo
  {pages} {L99} (\bibinfo {year} {2000})}\BibitemShut {NoStop}%
\bibitem [{\citenamefont {Eisenstein}\ and\ \citenamefont
  {Loeb}(1997)}]{eisenstein1997can}%
  \BibitemOpen
  \bibfield  {author} {\bibinfo {author} {\bibfnamefont {D.~J.}\ \bibnamefont
  {Eisenstein}}\ and\ \bibinfo {author} {\bibfnamefont {A.}~\bibnamefont
  {Loeb}},\ }\bibfield  {title} {\bibinfo {title} {Can the tully-fisher
  relation be the result of initial conditions?},\ }in\ \href@noop {} {\emph
  {\bibinfo {booktitle} {Galaxy Scaling Relations: Origins, Evolution and
  Applications}}}\ (\bibinfo  {publisher} {Springer},\ \bibinfo {year} {1997})\
  pp.\ \bibinfo {pages} {15--24}\BibitemShut {NoStop}%
\bibitem [{\citenamefont {Foot}\ and\ \citenamefont
  {Vagnozzi}(2015)}]{foot2015dissipative}%
  \BibitemOpen
  \bibfield  {author} {\bibinfo {author} {\bibfnamefont {R.}~\bibnamefont
  {Foot}}\ and\ \bibinfo {author} {\bibfnamefont {S.}~\bibnamefont
  {Vagnozzi}},\ }\bibfield  {title} {\bibinfo {title} {Dissipative hidden
  sector dark matter},\ }\href@noop {} {\bibfield  {journal} {\bibinfo
  {journal} {Physical Review D}\ }\textbf {\bibinfo {volume} {91}},\ \bibinfo
  {pages} {023512} (\bibinfo {year} {2015})}\BibitemShut {NoStop}%
\bibitem [{\citenamefont {Ackerman}\ \emph {et~al.}(2009)\citenamefont
  {Ackerman}, \citenamefont {Buckley}, \citenamefont {Carroll},\ and\
  \citenamefont {Kamionkowski}}]{ackerman2009dark}%
  \BibitemOpen
  \bibfield  {author} {\bibinfo {author} {\bibfnamefont {L.}~\bibnamefont
  {Ackerman}}, \bibinfo {author} {\bibfnamefont {M.~R.}\ \bibnamefont
  {Buckley}}, \bibinfo {author} {\bibfnamefont {S.~M.}\ \bibnamefont
  {Carroll}},\ and\ \bibinfo {author} {\bibfnamefont {M.}~\bibnamefont
  {Kamionkowski}},\ }\bibfield  {title} {\bibinfo {title} {Dark matter and dark
  radiation},\ }\href@noop {} {\bibfield  {journal} {\bibinfo  {journal}
  {Physical Review D}\ }\textbf {\bibinfo {volume} {79}},\ \bibinfo {pages}
  {023519} (\bibinfo {year} {2009})}\BibitemShut {NoStop}%
\bibitem [{\citenamefont {De~R{\'u}jula}\ \emph {et~al.}(1990)\citenamefont
  {De~R{\'u}jula}, \citenamefont {Glashow},\ and\ \citenamefont
  {Sarid}}]{de1990charged}%
  \BibitemOpen
  \bibfield  {author} {\bibinfo {author} {\bibfnamefont {A.}~\bibnamefont
  {De~R{\'u}jula}}, \bibinfo {author} {\bibfnamefont {S.}~\bibnamefont
  {Glashow}},\ and\ \bibinfo {author} {\bibfnamefont {U.}~\bibnamefont
  {Sarid}},\ }\bibfield  {title} {\bibinfo {title} {Charged dark matter},\
  }\href@noop {} {\bibfield  {journal} {\bibinfo  {journal} {Nuclear Physics
  B}\ }\textbf {\bibinfo {volume} {333}},\ \bibinfo {pages} {173} (\bibinfo
  {year} {1990})}\BibitemShut {NoStop}%
\bibitem [{\citenamefont {Mu{\~n}oz}\ and\ \citenamefont
  {Loeb}(2018)}]{munoz2018small}%
  \BibitemOpen
  \bibfield  {author} {\bibinfo {author} {\bibfnamefont {J.~B.}\ \bibnamefont
  {Mu{\~n}oz}}\ and\ \bibinfo {author} {\bibfnamefont {A.}~\bibnamefont
  {Loeb}},\ }\bibfield  {title} {\bibinfo {title} {A small amount of
  mini-charged dark matter could cool the baryons in the early universe},\
  }\href@noop {} {\bibfield  {journal} {\bibinfo  {journal} {Nature}\ }\textbf
  {\bibinfo {volume} {557}},\ \bibinfo {pages} {684} (\bibinfo {year}
  {2018})}\BibitemShut {NoStop}%
\bibitem [{\citenamefont {Agrawal}\ \emph {et~al.}(2017)\citenamefont
  {Agrawal}, \citenamefont {Cyr-Racine}, \citenamefont {Randall},\ and\
  \citenamefont {Scholtz}}]{agrawal2017make}%
  \BibitemOpen
  \bibfield  {author} {\bibinfo {author} {\bibfnamefont {P.}~\bibnamefont
  {Agrawal}}, \bibinfo {author} {\bibfnamefont {F.-Y.}\ \bibnamefont
  {Cyr-Racine}}, \bibinfo {author} {\bibfnamefont {L.}~\bibnamefont
  {Randall}},\ and\ \bibinfo {author} {\bibfnamefont {J.}~\bibnamefont
  {Scholtz}},\ }\bibfield  {title} {\bibinfo {title} {Make dark matter charged
  again},\ }\href@noop {} {\bibfield  {journal} {\bibinfo  {journal} {Journal
  of Cosmology and Astroparticle Physics}\ }\textbf {\bibinfo {volume}
  {2017}}\bibinfo  {number} { (05)},\ \bibinfo {pages} {022}}\BibitemShut
  {NoStop}%
\bibitem [{\citenamefont {Kamada}\ \emph {et~al.}(2020)\citenamefont {Kamada},
  \citenamefont {Yamada},\ and\ \citenamefont
  {Yanagida}}]{kamada2020unification}%
  \BibitemOpen
\bibfield  {number} {  }\bibfield  {author} {\bibinfo {author} {\bibfnamefont
  {A.}~\bibnamefont {Kamada}}, \bibinfo {author} {\bibfnamefont
  {M.}~\bibnamefont {Yamada}},\ and\ \bibinfo {author} {\bibfnamefont {T.~T.}\
  \bibnamefont {Yanagida}},\ }\bibfield  {title} {\bibinfo {title} {Unification
  for darkly charged dark matter},\ }\href@noop {} {\bibfield  {journal}
  {\bibinfo  {journal} {Physical Review D}\ }\textbf {\bibinfo {volume}
  {102}},\ \bibinfo {pages} {015012} (\bibinfo {year} {2020})}\BibitemShut
  {NoStop}%
\bibitem [{\citenamefont {Kramer}\ and\ \citenamefont
  {Randall}(2016)}]{kramer2016updated}%
  \BibitemOpen
  \bibfield  {author} {\bibinfo {author} {\bibfnamefont {E.~D.}\ \bibnamefont
  {Kramer}}\ and\ \bibinfo {author} {\bibfnamefont {L.}~\bibnamefont
  {Randall}},\ }\bibfield  {title} {\bibinfo {title} {Updated kinematic
  constraints on a dark disk},\ }\href@noop {} {\bibfield  {journal} {\bibinfo
  {journal} {The Astrophysical Journal}\ }\textbf {\bibinfo {volume} {824}},\
  \bibinfo {pages} {116} (\bibinfo {year} {2016})}\BibitemShut {NoStop}%
\bibitem [{\citenamefont {Randall}\ and\ \citenamefont
  {Reece}(2014)}]{randall2014dark}%
  \BibitemOpen
  \bibfield  {author} {\bibinfo {author} {\bibfnamefont {L.}~\bibnamefont
  {Randall}}\ and\ \bibinfo {author} {\bibfnamefont {M.}~\bibnamefont
  {Reece}},\ }\bibfield  {title} {\bibinfo {title} {Dark matter as a trigger
  for periodic comet impacts},\ }\href@noop {} {\bibfield  {journal} {\bibinfo
  {journal} {Physical review letters}\ }\textbf {\bibinfo {volume} {112}},\
  \bibinfo {pages} {161301} (\bibinfo {year} {2014})}\BibitemShut {NoStop}%
\bibitem [{\citenamefont {Fan}\ \emph {et~al.}(2013)\citenamefont {Fan},
  \citenamefont {Katz}, \citenamefont {Randall},\ and\ \citenamefont
  {Reece}}]{fan2013double}%
  \BibitemOpen
  \bibfield  {author} {\bibinfo {author} {\bibfnamefont {J.}~\bibnamefont
  {Fan}}, \bibinfo {author} {\bibfnamefont {A.}~\bibnamefont {Katz}}, \bibinfo
  {author} {\bibfnamefont {L.}~\bibnamefont {Randall}},\ and\ \bibinfo {author}
  {\bibfnamefont {M.}~\bibnamefont {Reece}},\ }\bibfield  {title} {\bibinfo
  {title} {Double-disk dark matter},\ }\href@noop {} {\bibfield  {journal}
  {\bibinfo  {journal} {Physics of the Dark Universe}\ }\textbf {\bibinfo
  {volume} {2}},\ \bibinfo {pages} {139} (\bibinfo {year} {2013})}\BibitemShut
  {NoStop}%
\bibitem [{\citenamefont {Keeton}(2001)}]{keeton2001catalog}%
  \BibitemOpen
  \bibfield  {author} {\bibinfo {author} {\bibfnamefont {C.~R.}\ \bibnamefont
  {Keeton}},\ }\bibfield  {title} {\bibinfo {title} {A catalog of mass models
  for gravitational lensing},\ }\href@noop {} {\bibfield  {journal} {\bibinfo
  {journal} {arXiv preprint astro-ph/0102341}\ } (\bibinfo {year}
  {2001})}\BibitemShut {NoStop}%
\bibitem [{\citenamefont {Martel}\ and\ \citenamefont
  {Shapiro}(2003)}]{martel2003gravitational}%
  \BibitemOpen
  \bibfield  {author} {\bibinfo {author} {\bibfnamefont {H.}~\bibnamefont
  {Martel}}\ and\ \bibinfo {author} {\bibfnamefont {P.~R.}\ \bibnamefont
  {Shapiro}},\ }\bibfield  {title} {\bibinfo {title} {Gravitational lensing by
  cdm halos: singular versus nonsingular profiles},\ }\href@noop {} {\bibfield
  {journal} {\bibinfo  {journal} {arXiv preprint astro-ph/0305174}\ } (\bibinfo
  {year} {2003})}\BibitemShut {NoStop}%
\bibitem [{\citenamefont {Persic}\ \emph {et~al.}(1996)\citenamefont {Persic},
  \citenamefont {Salucci},\ and\ \citenamefont {Stel}}]{persic1996universal}%
  \BibitemOpen
  \bibfield  {author} {\bibinfo {author} {\bibfnamefont {M.}~\bibnamefont
  {Persic}}, \bibinfo {author} {\bibfnamefont {P.}~\bibnamefont {Salucci}},\
  and\ \bibinfo {author} {\bibfnamefont {F.}~\bibnamefont {Stel}},\ }\bibfield
  {title} {\bibinfo {title} {The universal rotation curve of spiral galaxies-i.
  the dark matter connection},\ }\href@noop {} {\bibfield  {journal} {\bibinfo
  {journal} {Monthly Notices of the Royal Astronomical Society}\ }\textbf
  {\bibinfo {volume} {281}},\ \bibinfo {pages} {27} (\bibinfo {year}
  {1996})}\BibitemShut {NoStop}%
\bibitem [{\citenamefont {Navarro}\ \emph {et~al.}(1997)\citenamefont
  {Navarro}, \citenamefont {Frenk},\ and\ \citenamefont
  {White}}]{navarro1997universal}%
  \BibitemOpen
  \bibfield  {author} {\bibinfo {author} {\bibfnamefont {J.~F.}\ \bibnamefont
  {Navarro}}, \bibinfo {author} {\bibfnamefont {C.~S.}\ \bibnamefont {Frenk}},\
  and\ \bibinfo {author} {\bibfnamefont {S.~D.}\ \bibnamefont {White}},\
  }\bibfield  {title} {\bibinfo {title} {A universal density profile from
  hierarchical clustering},\ }\href@noop {} {\bibfield  {journal} {\bibinfo
  {journal} {The Astrophysical Journal}\ }\textbf {\bibinfo {volume} {490}},\
  \bibinfo {pages} {493} (\bibinfo {year} {1997})}\BibitemShut {NoStop}%
\bibitem [{\citenamefont {Carroll}(2001)}]{carroll2001cosmological}%
  \BibitemOpen
  \bibfield  {author} {\bibinfo {author} {\bibfnamefont {S.~M.}\ \bibnamefont
  {Carroll}},\ }\bibfield  {title} {\bibinfo {title} {The cosmological
  constant},\ }\href@noop {} {\bibfield  {journal} {\bibinfo  {journal} {Living
  Reviews in Relativity}\ }\textbf {\bibinfo {volume} {4}},\ \bibinfo {pages}
  {1} (\bibinfo {year} {2001})}\BibitemShut {NoStop}%
\bibitem [{\citenamefont {Bekenstein}(2004)}]{bekenstein2004relativistic}%
  \BibitemOpen
  \bibfield  {author} {\bibinfo {author} {\bibfnamefont {J.~D.}\ \bibnamefont
  {Bekenstein}},\ }\bibfield  {title} {\bibinfo {title} {Relativistic
  gravitation theory for the modified newtonian dynamics paradigm},\
  }\href@noop {} {\bibfield  {journal} {\bibinfo  {journal} {Physical Review
  D}\ }\textbf {\bibinfo {volume} {70}},\ \bibinfo {pages} {083509} (\bibinfo
  {year} {2004})}\BibitemShut {NoStop}%
\bibitem [{\citenamefont {McGaugh}(2015)}]{mcgaugh2015tale}%
  \BibitemOpen
  \bibfield  {author} {\bibinfo {author} {\bibfnamefont {S.~S.}\ \bibnamefont
  {McGaugh}},\ }\bibfield  {title} {\bibinfo {title} {A tale of two paradigms:
  the mutual incommensurability of $\lambda$cdm and mond},\ }\href@noop {}
  {\bibfield  {journal} {\bibinfo  {journal} {Canadian Journal of Physics}\
  }\textbf {\bibinfo {volume} {93}},\ \bibinfo {pages} {250} (\bibinfo {year}
  {2015})}\BibitemShut {NoStop}%
\bibitem [{\citenamefont {McGaugh}(2005)}]{mcgaugh2005balance}%
  \BibitemOpen
  \bibfield  {author} {\bibinfo {author} {\bibfnamefont {S.~S.}\ \bibnamefont
  {McGaugh}},\ }\bibfield  {title} {\bibinfo {title} {Balance of dark and
  luminous mass in rotating galaxies},\ }\href@noop {} {\bibfield  {journal}
  {\bibinfo  {journal} {Physical Review Letters}\ }\textbf {\bibinfo {volume}
  {95}},\ \bibinfo {pages} {171302} (\bibinfo {year} {2005})}\BibitemShut
  {NoStop}%
\bibitem [{\citenamefont {Kroupa}\ \emph {et~al.}(2012)\citenamefont {Kroupa},
  \citenamefont {Pawlowski},\ and\ \citenamefont
  {Milgrom}}]{kroupa2012failures}%
  \BibitemOpen
  \bibfield  {author} {\bibinfo {author} {\bibfnamefont {P.}~\bibnamefont
  {Kroupa}}, \bibinfo {author} {\bibfnamefont {M.}~\bibnamefont {Pawlowski}},\
  and\ \bibinfo {author} {\bibfnamefont {M.}~\bibnamefont {Milgrom}},\
  }\bibfield  {title} {\bibinfo {title} {The failures of the standard model of
  cosmology require a new paradigm},\ }\href@noop {} {\bibfield  {journal}
  {\bibinfo  {journal} {International Journal of Modern Physics D}\ }\textbf
  {\bibinfo {volume} {21}},\ \bibinfo {pages} {1230003} (\bibinfo {year}
  {2012})}\BibitemShut {NoStop}%
\bibitem [{\citenamefont {Clifton}\ \emph {et~al.}(2012)\citenamefont
  {Clifton}, \citenamefont {Ferreira}, \citenamefont {Padilla},\ and\
  \citenamefont {Skordis}}]{clifton2012modified}%
  \BibitemOpen
  \bibfield  {author} {\bibinfo {author} {\bibfnamefont {T.}~\bibnamefont
  {Clifton}}, \bibinfo {author} {\bibfnamefont {P.~G.}\ \bibnamefont
  {Ferreira}}, \bibinfo {author} {\bibfnamefont {A.}~\bibnamefont {Padilla}},\
  and\ \bibinfo {author} {\bibfnamefont {C.}~\bibnamefont {Skordis}},\
  }\bibfield  {title} {\bibinfo {title} {Modified gravity and cosmology},\
  }\href@noop {} {\bibfield  {journal} {\bibinfo  {journal} {Physics Reports}\
  }\textbf {\bibinfo {volume} {513}},\ \bibinfo {pages} {1} (\bibinfo {year}
  {2012})}\BibitemShut {NoStop}%
\bibitem [{\citenamefont {Rahvar}\ and\ \citenamefont
  {Mashhoon}(2014)}]{rahvar2014observational}%
  \BibitemOpen
  \bibfield  {author} {\bibinfo {author} {\bibfnamefont {S.}~\bibnamefont
  {Rahvar}}\ and\ \bibinfo {author} {\bibfnamefont {B.}~\bibnamefont
  {Mashhoon}},\ }\bibfield  {title} {\bibinfo {title} {Observational tests of
  nonlocal gravity: galaxy rotation curves and clusters of galaxies},\
  }\href@noop {} {\bibfield  {journal} {\bibinfo  {journal} {Physical Review
  D}\ }\textbf {\bibinfo {volume} {89}},\ \bibinfo {pages} {104011} (\bibinfo
  {year} {2014})}\BibitemShut {NoStop}%
\bibitem [{\citenamefont {Hossenfelder}\ and\ \citenamefont
  {Mistele}(2019)}]{hossenfelder2019strong}%
  \BibitemOpen
  \bibfield  {author} {\bibinfo {author} {\bibfnamefont {S.}~\bibnamefont
  {Hossenfelder}}\ and\ \bibinfo {author} {\bibfnamefont {T.}~\bibnamefont
  {Mistele}},\ }\bibfield  {title} {\bibinfo {title} {Strong lensing with
  superfluid dark matter},\ }\href@noop {} {\bibfield  {journal} {\bibinfo
  {journal} {Journal of Cosmology and Astroparticle Physics}\ }\textbf
  {\bibinfo {volume} {2019}}\bibinfo  {number} { (02)},\ \bibinfo {pages}
  {001}}\BibitemShut {NoStop}%
\bibitem [{\citenamefont {McGaugh}(2012)}]{mcgaugh2012baryonic}%
  \BibitemOpen
\bibfield  {number} {  }\bibfield  {author} {\bibinfo {author} {\bibfnamefont
  {S.~S.}\ \bibnamefont {McGaugh}},\ }\bibfield  {title} {\bibinfo {title} {The
  baryonic tully-fisher relation of gas-rich galaxies as a test of $\lambda$cdm
  and mond},\ }\href@noop {} {\bibfield  {journal} {\bibinfo  {journal} {The
  Astronomical Journal}\ }\textbf {\bibinfo {volume} {143}},\ \bibinfo {pages}
  {40} (\bibinfo {year} {2012})}\BibitemShut {NoStop}%
\bibitem [{\citenamefont {Sumner}(2002)}]{sumner2002experimental}%
  \BibitemOpen
  \bibfield  {author} {\bibinfo {author} {\bibfnamefont {T.~J.}\ \bibnamefont
  {Sumner}},\ }\bibfield  {title} {\bibinfo {title} {Experimental searches for
  dark matter},\ }\href@noop {} {\bibfield  {journal} {\bibinfo  {journal}
  {Living Reviews in Relativity}\ }\textbf {\bibinfo {volume} {5}},\ \bibinfo
  {pages} {4} (\bibinfo {year} {2002})}\BibitemShut {NoStop}%
\bibitem [{\citenamefont {Ishak}(2019)}]{ishak2019testing}%
  \BibitemOpen
  \bibfield  {author} {\bibinfo {author} {\bibfnamefont {M.}~\bibnamefont
  {Ishak}},\ }\bibfield  {title} {\bibinfo {title} {Testing general relativity
  in cosmology},\ }\href@noop {} {\bibfield  {journal} {\bibinfo  {journal}
  {Living Reviews in Relativity}\ }\textbf {\bibinfo {volume} {22}},\ \bibinfo
  {pages} {1} (\bibinfo {year} {2019})}\BibitemShut {NoStop}%
\bibitem [{\citenamefont {Cveti{\v{c}}}\ \emph {et~al.}(2002)\citenamefont
  {Cveti{\v{c}}}, \citenamefont {Gibbons}, \citenamefont {L{\"u}},\ and\
  \citenamefont {Pope}}]{cvetivc2002m}%
  \BibitemOpen
  \bibfield  {author} {\bibinfo {author} {\bibfnamefont {M.}~\bibnamefont
  {Cveti{\v{c}}}}, \bibinfo {author} {\bibfnamefont {G.}~\bibnamefont
  {Gibbons}}, \bibinfo {author} {\bibfnamefont {H.}~\bibnamefont {L{\"u}}},\
  and\ \bibinfo {author} {\bibfnamefont {C.}~\bibnamefont {Pope}},\ }\bibfield
  {title} {\bibinfo {title} {M-theory conifolds},\ }\href@noop {} {\bibfield
  {journal} {\bibinfo  {journal} {Physical Review Letters}\ }\textbf {\bibinfo
  {volume} {88}},\ \bibinfo {pages} {121602} (\bibinfo {year}
  {2002})}\BibitemShut {NoStop}%
\bibitem [{\citenamefont {Cveti{\v{c}}}\ and\ \citenamefont
  {Liu}(2005)}]{cvetivc2005supersymmetric}%
  \BibitemOpen
  \bibfield  {author} {\bibinfo {author} {\bibfnamefont {M.}~\bibnamefont
  {Cveti{\v{c}}}}\ and\ \bibinfo {author} {\bibfnamefont {T.}~\bibnamefont
  {Liu}},\ }\bibfield  {title} {\bibinfo {title} {Supersymmetric standard
  models, flux compactification and moduli stabilization},\ }\href@noop {}
  {\bibfield  {journal} {\bibinfo  {journal} {Physics Letters B}\ }\textbf
  {\bibinfo {volume} {610}},\ \bibinfo {pages} {122} (\bibinfo {year}
  {2005})}\BibitemShut {NoStop}%
\bibitem [{\citenamefont {Cveti{\v{c}}}\ \emph {et~al.}(2005)\citenamefont
  {Cveti{\v{c}}}, \citenamefont {L{\"u}}, \citenamefont {Page},\ and\
  \citenamefont {Pope}}]{cvetivc2005new}%
  \BibitemOpen
  \bibfield  {author} {\bibinfo {author} {\bibfnamefont {M.}~\bibnamefont
  {Cveti{\v{c}}}}, \bibinfo {author} {\bibfnamefont {H.}~\bibnamefont
  {L{\"u}}}, \bibinfo {author} {\bibfnamefont {D.~N.}\ \bibnamefont {Page}},\
  and\ \bibinfo {author} {\bibfnamefont {C.}~\bibnamefont {Pope}},\ }\bibfield
  {title} {\bibinfo {title} {New einstein-sasaki spaces in five and higher
  dimensions},\ }\href@noop {} {\bibfield  {journal} {\bibinfo  {journal}
  {Physical Review Letters}\ }\textbf {\bibinfo {volume} {95}},\ \bibinfo
  {pages} {071101} (\bibinfo {year} {2005})}\BibitemShut {NoStop}%
\bibitem [{\citenamefont {Catena}\ and\ \citenamefont
  {Covi}(2014)}]{catena2014susy}%
  \BibitemOpen
  \bibfield  {author} {\bibinfo {author} {\bibfnamefont {R.}~\bibnamefont
  {Catena}}\ and\ \bibinfo {author} {\bibfnamefont {L.}~\bibnamefont {Covi}},\
  }\bibfield  {title} {\bibinfo {title} {Susy dark matter (s)},\ }in\
  \href@noop {} {\emph {\bibinfo {booktitle} {Supersymmetry After the Higgs
  Discovery}}}\ (\bibinfo  {publisher} {Springer},\ \bibinfo {year} {2014})\
  pp.\ \bibinfo {pages} {121--136}\BibitemShut {NoStop}%
\bibitem [{\citenamefont {Tamaki}(2008)}]{tamaki2008post}%
  \BibitemOpen
  \bibfield  {author} {\bibinfo {author} {\bibfnamefont {T.}~\bibnamefont
  {Tamaki}},\ }\bibfield  {title} {\bibinfo {title} {Post-newtonian parameters
  in the tensor-vector-scalar theory},\ }\href@noop {} {\bibfield  {journal}
  {\bibinfo  {journal} {Physical Review D}\ }\textbf {\bibinfo {volume} {77}},\
  \bibinfo {pages} {124020} (\bibinfo {year} {2008})}\BibitemShut {NoStop}%
\bibitem [{\citenamefont {Konno}\ \emph {et~al.}(2008)\citenamefont {Konno},
  \citenamefont {Matsuyama}, \citenamefont {Asano},\ and\ \citenamefont
  {Tanda}}]{konno2008flat}%
  \BibitemOpen
  \bibfield  {author} {\bibinfo {author} {\bibfnamefont {K.}~\bibnamefont
  {Konno}}, \bibinfo {author} {\bibfnamefont {T.}~\bibnamefont {Matsuyama}},
  \bibinfo {author} {\bibfnamefont {Y.}~\bibnamefont {Asano}},\ and\ \bibinfo
  {author} {\bibfnamefont {S.}~\bibnamefont {Tanda}},\ }\bibfield  {title}
  {\bibinfo {title} {Flat rotation curves in chern-simons modified gravity},\
  }\href@noop {} {\bibfield  {journal} {\bibinfo  {journal} {Physical Review
  D}\ }\textbf {\bibinfo {volume} {78}},\ \bibinfo {pages} {024037} (\bibinfo
  {year} {2008})}\BibitemShut {NoStop}%
\bibitem [{\citenamefont {Sotani}(2010)}]{sotani2010slowly}%
  \BibitemOpen
  \bibfield  {author} {\bibinfo {author} {\bibfnamefont {H.}~\bibnamefont
  {Sotani}},\ }\bibfield  {title} {\bibinfo {title} {Slowly rotating
  relativistic stars in tensor-vector-scalar theory},\ }\href@noop {}
  {\bibfield  {journal} {\bibinfo  {journal} {Physical Review D}\ }\textbf
  {\bibinfo {volume} {81}},\ \bibinfo {pages} {084006} (\bibinfo {year}
  {2010})}\BibitemShut {NoStop}%
\bibitem [{\citenamefont {Stabile}\ and\ \citenamefont
  {Scelza}(2011)}]{stabile2011rotation}%
  \BibitemOpen
  \bibfield  {author} {\bibinfo {author} {\bibfnamefont {A.}~\bibnamefont
  {Stabile}}\ and\ \bibinfo {author} {\bibfnamefont {G.}~\bibnamefont
  {Scelza}},\ }\bibfield  {title} {\bibinfo {title} {Rotation curves of
  galaxies by fourth order gravity},\ }\href@noop {} {\bibfield  {journal}
  {\bibinfo  {journal} {Physical Review D}\ }\textbf {\bibinfo {volume} {84}},\
  \bibinfo {pages} {124023} (\bibinfo {year} {2011})}\BibitemShut {NoStop}%
\bibitem [{\citenamefont {Stabile}\ and\ \citenamefont
  {Capozziello}(2013)}]{stabile2013galaxy}%
  \BibitemOpen
  \bibfield  {author} {\bibinfo {author} {\bibfnamefont {A.}~\bibnamefont
  {Stabile}}\ and\ \bibinfo {author} {\bibfnamefont {S.}~\bibnamefont
  {Capozziello}},\ }\bibfield  {title} {\bibinfo {title} {Galaxy rotation
  curves in f (r, $\phi$) gravity},\ }\href@noop {} {\bibfield  {journal}
  {\bibinfo  {journal} {Physical Review D}\ }\textbf {\bibinfo {volume} {87}},\
  \bibinfo {pages} {064002} (\bibinfo {year} {2013})}\BibitemShut {NoStop}%
\bibitem [{\citenamefont {Diez-Tejedor}\ \emph {et~al.}(2018)\citenamefont
  {Diez-Tejedor}, \citenamefont {Flores},\ and\ \citenamefont
  {Niz}}]{diez2018horndeski}%
  \BibitemOpen
  \bibfield  {author} {\bibinfo {author} {\bibfnamefont {A.}~\bibnamefont
  {Diez-Tejedor}}, \bibinfo {author} {\bibfnamefont {F.}~\bibnamefont
  {Flores}},\ and\ \bibinfo {author} {\bibfnamefont {G.}~\bibnamefont {Niz}},\
  }\bibfield  {title} {\bibinfo {title} {Horndeski dark matter and beyond},\
  }\href@noop {} {\bibfield  {journal} {\bibinfo  {journal} {Physical Review
  D}\ }\textbf {\bibinfo {volume} {97}},\ \bibinfo {pages} {123524} (\bibinfo
  {year} {2018})}\BibitemShut {NoStop}%
\bibitem [{\citenamefont {Berezhiani}\ and\ \citenamefont
  {Khoury}(2015)}]{berezhiani2015theory}%
  \BibitemOpen
  \bibfield  {author} {\bibinfo {author} {\bibfnamefont {L.}~\bibnamefont
  {Berezhiani}}\ and\ \bibinfo {author} {\bibfnamefont {J.}~\bibnamefont
  {Khoury}},\ }\bibfield  {title} {\bibinfo {title} {Theory of dark matter
  superfluidity},\ }\href@noop {} {\bibfield  {journal} {\bibinfo  {journal}
  {Physical Review D}\ }\textbf {\bibinfo {volume} {92}},\ \bibinfo {pages}
  {103510} (\bibinfo {year} {2015})}\BibitemShut {NoStop}%
\bibitem [{\citenamefont {Sharma}\ \emph {et~al.}(2019)\citenamefont {Sharma},
  \citenamefont {Khoury},\ and\ \citenamefont {Lubensky}}]{sharma2019equation}%
  \BibitemOpen
  \bibfield  {author} {\bibinfo {author} {\bibfnamefont {A.}~\bibnamefont
  {Sharma}}, \bibinfo {author} {\bibfnamefont {J.}~\bibnamefont {Khoury}},\
  and\ \bibinfo {author} {\bibfnamefont {T.}~\bibnamefont {Lubensky}},\
  }\bibfield  {title} {\bibinfo {title} {The equation of state of dark matter
  superfluids},\ }\href@noop {} {\bibfield  {journal} {\bibinfo  {journal}
  {Journal of Cosmology and Astroparticle Physics}\ }\textbf {\bibinfo {volume}
  {2019}}\bibinfo  {number} { (05)},\ \bibinfo {pages} {054}}\BibitemShut
  {NoStop}%
\bibitem [{\citenamefont {Verlinde}(2017)}]{verlinde2017emergent}%
  \BibitemOpen
\bibfield  {number} {  }\bibfield  {author} {\bibinfo {author} {\bibfnamefont
  {E.~P.}\ \bibnamefont {Verlinde}},\ }\bibfield  {title} {\bibinfo {title}
  {{Emergent Gravity and the Dark Universe}},\ }\href@noop {} {\bibfield
  {journal} {\bibinfo  {journal} {SciPost Physics}\ }\textbf {\bibinfo {volume}
  {2}},\ \bibinfo {pages} {016} (\bibinfo {year} {2017})}\BibitemShut {NoStop}%
\bibitem [{\citenamefont {Verlinde}(2011)}]{verlinde2011origin}%
  \BibitemOpen
  \bibfield  {author} {\bibinfo {author} {\bibfnamefont {E.}~\bibnamefont
  {Verlinde}},\ }\bibfield  {title} {\bibinfo {title} {On the origin of gravity
  and the laws of newton},\ }\href@noop {} {\bibfield  {journal} {\bibinfo
  {journal} {Journal of High Energy Physics}\ }\textbf {\bibinfo {volume}
  {2011}},\ \bibinfo {pages} {1} (\bibinfo {year} {2011})}\BibitemShut
  {NoStop}%
\bibitem [{\citenamefont {Brouwer}\ \emph {et~al.}(2017)\citenamefont
  {Brouwer}, \citenamefont {Visser}, \citenamefont {Dvornik}, \citenamefont
  {Hoekstra}, \citenamefont {Kuijken}, \citenamefont {Valentijn}, \citenamefont
  {Bilicki}, \citenamefont {Blake}, \citenamefont {Brough}, \citenamefont
  {Buddelmeijer} \emph {et~al.}}]{brouwer2017first}%
  \BibitemOpen
  \bibfield  {author} {\bibinfo {author} {\bibfnamefont {M.~M.}\ \bibnamefont
  {Brouwer}}, \bibinfo {author} {\bibfnamefont {M.~R.}\ \bibnamefont {Visser}},
  \bibinfo {author} {\bibfnamefont {A.}~\bibnamefont {Dvornik}}, \bibinfo
  {author} {\bibfnamefont {H.}~\bibnamefont {Hoekstra}}, \bibinfo {author}
  {\bibfnamefont {K.}~\bibnamefont {Kuijken}}, \bibinfo {author} {\bibfnamefont
  {E.~A.}\ \bibnamefont {Valentijn}}, \bibinfo {author} {\bibfnamefont
  {M.}~\bibnamefont {Bilicki}}, \bibinfo {author} {\bibfnamefont
  {C.}~\bibnamefont {Blake}}, \bibinfo {author} {\bibfnamefont
  {S.}~\bibnamefont {Brough}}, \bibinfo {author} {\bibfnamefont
  {H.}~\bibnamefont {Buddelmeijer}}, \emph {et~al.},\ }\bibfield  {title}
  {\bibinfo {title} {First test of verlinde's theory of emergent gravity using
  weak gravitational lensing measurements},\ }\href@noop {} {\bibfield
  {journal} {\bibinfo  {journal} {Monthly Notices of the Royal Astronomical
  Society}\ }\textbf {\bibinfo {volume} {466}},\ \bibinfo {pages} {2547}
  (\bibinfo {year} {2017})}\BibitemShut {NoStop}%
\bibitem [{\citenamefont {Huebener}(2001)}]{huebener2001ginzburg}%
  \BibitemOpen
  \bibfield  {author} {\bibinfo {author} {\bibfnamefont {R.~P.}\ \bibnamefont
  {Huebener}},\ }\bibfield  {title} {\bibinfo {title} {Ginzburg-landau
  theory},\ }in\ \href@noop {} {\emph {\bibinfo {booktitle} {Magnetic Flux
  Structures in Superconductors}}}\ (\bibinfo  {publisher} {Springer},\
  \bibinfo {year} {2001})\ pp.\ \bibinfo {pages} {33--57}\BibitemShut {NoStop}%
\bibitem [{\citenamefont {Gross}(1961)}]{gross1961structure}%
  \BibitemOpen
  \bibfield  {author} {\bibinfo {author} {\bibfnamefont {E.~P.}\ \bibnamefont
  {Gross}},\ }\bibfield  {title} {\bibinfo {title} {Structure of a quantized
  vortex in boson systems},\ }\href@noop {} {\bibfield  {journal} {\bibinfo
  {journal} {Il Nuovo Cimento (1955-1965)}\ }\textbf {\bibinfo {volume} {20}},\
  \bibinfo {pages} {454} (\bibinfo {year} {1961})}\BibitemShut {NoStop}%
\bibitem [{\citenamefont {Pitaevskii}(1961)}]{pitaevskii1961vortex}%
  \BibitemOpen
  \bibfield  {author} {\bibinfo {author} {\bibfnamefont {L.~P.}\ \bibnamefont
  {Pitaevskii}},\ }\bibfield  {title} {\bibinfo {title} {Vortex lines in an
  imperfect bose gas},\ }\href@noop {} {\bibfield  {journal} {\bibinfo
  {journal} {Soviet Physics JETP}\ }\textbf {\bibinfo {volume} {13}},\ \bibinfo
  {pages} {451} (\bibinfo {year} {1961})}\BibitemShut {NoStop}%
\bibitem [{Note1()}]{Note1}%
  \BibitemOpen
  \bibinfo {note} {A $|\Psi |^4$ term in the Ginzburg-Landau (Gross-Pitaevskii)
  free energy arises from a two particle interaction, whereas a $|\Psi |^6$
  term arises from a three particle interaction. Superfluid dark matter
  theories correspond to a three-particle interaction term but with a negative
  sign\cite {berezhiani2015theory}}\BibitemShut {NoStop}%
\bibitem [{\citenamefont {Maldacena}(1999)}]{maldacena1999large}%
  \BibitemOpen
  \bibfield  {author} {\bibinfo {author} {\bibfnamefont {J.}~\bibnamefont
  {Maldacena}},\ }\bibfield  {title} {\bibinfo {title} {The large-n limit of
  superconformal field theories and supergravity},\ }\href@noop {} {\bibfield
  {journal} {\bibinfo  {journal} {International Journal of Theoretical
  Physics}\ }\textbf {\bibinfo {volume} {38}},\ \bibinfo {pages} {1113}
  (\bibinfo {year} {1999})}\BibitemShut {NoStop}%
\bibitem [{\citenamefont {Susskind}(1995)}]{susskind1995world}%
  \BibitemOpen
  \bibfield  {author} {\bibinfo {author} {\bibfnamefont {L.}~\bibnamefont
  {Susskind}},\ }\bibfield  {title} {\bibinfo {title} {The world as a
  hologram},\ }\href@noop {} {\bibfield  {journal} {\bibinfo  {journal}
  {Journal of Mathematical Physics}\ }\textbf {\bibinfo {volume} {36}},\
  \bibinfo {pages} {6377} (\bibinfo {year} {1995})}\BibitemShut {NoStop}%
\bibitem [{\citenamefont {Bousso}(2002)}]{bousso2002holographic}%
  \BibitemOpen
  \bibfield  {author} {\bibinfo {author} {\bibfnamefont {R.}~\bibnamefont
  {Bousso}},\ }\bibfield  {title} {\bibinfo {title} {The holographic
  principle},\ }\href@noop {} {\bibfield  {journal} {\bibinfo  {journal}
  {Reviews of Modern Physics}\ }\textbf {\bibinfo {volume} {74}},\ \bibinfo
  {pages} {825} (\bibinfo {year} {2002})}\BibitemShut {NoStop}%
\bibitem [{\citenamefont {Hawking}\ and\ \citenamefont
  {Ellis}(1973)}]{hawking1973large}%
  \BibitemOpen
  \bibfield  {author} {\bibinfo {author} {\bibfnamefont {S.~W.}\ \bibnamefont
  {Hawking}}\ and\ \bibinfo {author} {\bibfnamefont {G.~F.~R.}\ \bibnamefont
  {Ellis}},\ }\href@noop {} {\emph {\bibinfo {title} {The large scale structure
  of space-time}}},\ Vol.~\bibinfo {volume} {1}\ (\bibinfo  {publisher}
  {Cambridge university press},\ \bibinfo {year} {1973})\BibitemShut {NoStop}%
\bibitem [{\citenamefont {Einstein}(1997)}]{einstein1997volume}%
  \BibitemOpen
  \bibfield  {author} {\bibinfo {author} {\bibfnamefont {A.}~\bibnamefont
  {Einstein}},\ }\href@noop {} {\emph {\bibinfo {title} {Volume 6. The Berlin
  Years: Writings, 1914-1917: English Translation of Selected Texts}}}\
  (\bibinfo  {publisher} {University Press},\ \bibinfo {year}
  {1997})\BibitemShut {NoStop}%
\bibitem [{\citenamefont {Tauber}(1979)}]{tauber1979albert}%
  \BibitemOpen
  \bibfield  {author} {\bibinfo {author} {\bibfnamefont {G.~E.}\ \bibnamefont
  {Tauber}},\ }\href@noop {} {\emph {\bibinfo {title} {Albert Einstein's theory
  of general relativity}}}\ (\bibinfo  {publisher} {Crown Pub},\ \bibinfo
  {year} {1979})\BibitemShut {NoStop}%
\bibitem [{\citenamefont {Dirac}(1981)}]{dirac1981principles}%
  \BibitemOpen
  \bibfield  {author} {\bibinfo {author} {\bibfnamefont {P.~A.~M.}\
  \bibnamefont {Dirac}},\ }\href@noop {} {\emph {\bibinfo {title} {The
  principles of quantum mechanics}}}\ (\bibinfo  {publisher} {Oxford university
  press},\ \bibinfo {year} {1981})\BibitemShut {NoStop}%
\bibitem [{\citenamefont {Yang}\ and\ \citenamefont
  {Mills}(1954)}]{yang1954conservation}%
  \BibitemOpen
  \bibfield  {author} {\bibinfo {author} {\bibfnamefont {C.-N.}\ \bibnamefont
  {Yang}}\ and\ \bibinfo {author} {\bibfnamefont {R.~L.}\ \bibnamefont
  {Mills}},\ }\bibfield  {title} {\bibinfo {title} {Conservation of isotopic
  spin and isotopic gauge invariance},\ }\href@noop {} {\bibfield  {journal}
  {\bibinfo  {journal} {Physical Review}\ }\textbf {\bibinfo {volume} {96}},\
  \bibinfo {pages} {191} (\bibinfo {year} {1954})}\BibitemShut {NoStop}%
\bibitem [{\citenamefont {Koren}(2020)}]{koren2020hierarchy}%
  \BibitemOpen
  \bibfield  {author} {\bibinfo {author} {\bibfnamefont {S.}~\bibnamefont
  {Koren}},\ }\bibfield  {title} {\bibinfo {title} {The hierarchy problem: From
  the fundamentals to the frontiers},\ }\href@noop {} {\bibfield  {journal}
  {\bibinfo  {journal} {arXiv preprint arXiv:2009.11870}\ } (\bibinfo {year}
  {2020})}\BibitemShut {NoStop}%
\bibitem [{\citenamefont {Buras}\ \emph {et~al.}(1978)\citenamefont {Buras},
  \citenamefont {Ellis}, \citenamefont {Gaillard},\ and\ \citenamefont
  {Nanopoulos}}]{buras1978aspects}%
  \BibitemOpen
  \bibfield  {author} {\bibinfo {author} {\bibfnamefont {A.~J.}\ \bibnamefont
  {Buras}}, \bibinfo {author} {\bibfnamefont {J.}~\bibnamefont {Ellis}},
  \bibinfo {author} {\bibfnamefont {M.~K.}\ \bibnamefont {Gaillard}},\ and\
  \bibinfo {author} {\bibfnamefont {D.~V.}\ \bibnamefont {Nanopoulos}},\
  }\bibfield  {title} {\bibinfo {title} {Aspects of the grand unification of
  strong, weak and electromagnetic interactions},\ }\href@noop {} {\bibfield
  {journal} {\bibinfo  {journal} {Nuclear Physics B}\ }\textbf {\bibinfo
  {volume} {135}},\ \bibinfo {pages} {66} (\bibinfo {year} {1978})}\BibitemShut
  {NoStop}%
\bibitem [{Note2()}]{Note2}%
  \BibitemOpen
  \bibinfo {note} {At this scale, the strong, weak and electromagnetic forces
  are governed by a simple lie group such as SU($N$), with a single coupling
  $g$. This is implicit here.}\BibitemShut {Stop}%
\bibitem [{\citenamefont {Brown}(2005)}]{brown2005feynman}%
  \BibitemOpen
  \bibfield  {author} {\bibinfo {author} {\bibfnamefont {L.~M.}\ \bibnamefont
  {Brown}},\ }\href@noop {} {\emph {\bibinfo {title} {Feynman's Thesis-A New
  Approach to Quantum Theory}}}\ (\bibinfo  {publisher} {World Scientific
  Publishing},\ \bibinfo {year} {2005})\BibitemShut {NoStop}%
\bibitem [{\citenamefont {Fick}(1855)}]{fick1855v}%
  \BibitemOpen
  \bibfield  {author} {\bibinfo {author} {\bibfnamefont {A.}~\bibnamefont
  {Fick}},\ }\bibfield  {title} {\bibinfo {title} {V. on liquid diffusion},\
  }\href@noop {} {\bibfield  {journal} {\bibinfo  {journal} {The London,
  Edinburgh, and Dublin Philosophical Magazine and Journal of Science}\
  }\textbf {\bibinfo {volume} {10}},\ \bibinfo {pages} {30} (\bibinfo {year}
  {1855})}\BibitemShut {NoStop}%
\bibitem [{\citenamefont {Bingham}\ and\ \citenamefont
  {Mendon{\c{c}}a}(2007)}]{bingham2007can}%
  \BibitemOpen
  \bibfield  {author} {\bibinfo {author} {\bibfnamefont {R.}~\bibnamefont
  {Bingham}}\ and\ \bibinfo {author} {\bibfnamefont {J.~T.}\ \bibnamefont
  {Mendon{\c{c}}a}},\ }\bibfield  {title} {\bibinfo {title} {Can experiment
  access planck-scale physics?},\ }\href@noop {} {\bibfield  {journal}
  {\bibinfo  {journal} {AAPPS Bulletin}\ }\textbf {\bibinfo {volume} {17}},\
  \bibinfo {pages} {31} (\bibinfo {year} {2007})}\BibitemShut {NoStop}%
\bibitem [{Note3()}]{Note3}%
  \BibitemOpen
  \bibinfo {note} {Confer eq. (\ref {friction}) with $\protect \vec {v} = g\mu
  \protect \vec {E}$, where $\protect \vec {v}$ is the velocity and $\protect
  \vec {E}$ is the electric field}\BibitemShut {NoStop}%
\bibitem [{Note4()}]{Note4}%
  \BibitemOpen
  \bibinfo {note} {This is just the definition of an affine parameter. This is
  the unique transformation of the parameter $\tau $ which leaves the geodesic
  equation invariant.}\BibitemShut {Stop}%
\bibitem [{\citenamefont {Lemons}\ and\ \citenamefont
  {Gythiel}(1997)}]{lemons1997paul}%
  \BibitemOpen
  \bibfield  {author} {\bibinfo {author} {\bibfnamefont {D.~S.}\ \bibnamefont
  {Lemons}}\ and\ \bibinfo {author} {\bibfnamefont {A.}~\bibnamefont
  {Gythiel}},\ }\bibfield  {title} {\bibinfo {title} {Paul langevin’s 1908
  paper “on the theory of brownian motion”[“sur la th{\'e}orie du
  mouvement brownien,” cr acad. sci.(paris) 146, 530--533 (1908)]},\
  }\href@noop {} {\bibfield  {journal} {\bibinfo  {journal} {American Journal
  of Physics}\ }\textbf {\bibinfo {volume} {65}},\ \bibinfo {pages} {1079}
  (\bibinfo {year} {1997})}\BibitemShut {NoStop}%
\bibitem [{\citenamefont {Platen}(1986)}]{platen1986risken}%
  \BibitemOpen
  \bibfield  {author} {\bibinfo {author} {\bibfnamefont {E.}~\bibnamefont
  {Platen}},\ }\bibfield  {title} {\bibinfo {title} {Risken, h.: The
  fokker-planck equation. methods of solution and applications. springer series
  in synergetics, vol. 18. springer-verlag, berlin—heidelberg—new
  york—tokyo 1984, xvi, 454 pp., 95 figs., dm 125,-. isbn 3--540--13098--5},\
  }\href@noop {} {\bibfield  {journal} {\bibinfo  {journal} {Biometrical
  Journal}\ }\textbf {\bibinfo {volume} {28}},\ \bibinfo {pages} {740}
  (\bibinfo {year} {1986})}\BibitemShut {NoStop}%
\bibitem [{\citenamefont {Ross}\ \emph {et~al.}(1996)\citenamefont {Ross},
  \citenamefont {Kelly}, \citenamefont {Sullivan}, \citenamefont {Perry},
  \citenamefont {Mercer}, \citenamefont {Davis}, \citenamefont {Washburn},
  \citenamefont {Sager}, \citenamefont {Boyce},\ and\ \citenamefont
  {Bristow}}]{ross1996stochastic}%
  \BibitemOpen
  \bibfield  {author} {\bibinfo {author} {\bibfnamefont {S.~M.}\ \bibnamefont
  {Ross}}, \bibinfo {author} {\bibfnamefont {J.~J.}\ \bibnamefont {Kelly}},
  \bibinfo {author} {\bibfnamefont {R.~J.}\ \bibnamefont {Sullivan}}, \bibinfo
  {author} {\bibfnamefont {W.~J.}\ \bibnamefont {Perry}}, \bibinfo {author}
  {\bibfnamefont {D.}~\bibnamefont {Mercer}}, \bibinfo {author} {\bibfnamefont
  {R.~M.}\ \bibnamefont {Davis}}, \bibinfo {author} {\bibfnamefont {T.~D.}\
  \bibnamefont {Washburn}}, \bibinfo {author} {\bibfnamefont {E.~V.}\
  \bibnamefont {Sager}}, \bibinfo {author} {\bibfnamefont {J.~B.}\ \bibnamefont
  {Boyce}},\ and\ \bibinfo {author} {\bibfnamefont {V.~L.}\ \bibnamefont
  {Bristow}},\ }\href@noop {} {\emph {\bibinfo {title} {Stochastic
  processes}}},\ Vol.~\bibinfo {volume} {2}\ (\bibinfo  {publisher} {Wiley New
  York},\ \bibinfo {year} {1996})\BibitemShut {NoStop}%
\bibitem [{\citenamefont {Chan}(2013)}]{chan2013reconciliation}%
  \BibitemOpen
  \bibfield  {author} {\bibinfo {author} {\bibfnamefont {M.~H.}\ \bibnamefont
  {Chan}},\ }\bibfield  {title} {\bibinfo {title} {Reconciliation of modified
  newtonian dynamics and dark matter theory},\ }\href@noop {} {\bibfield
  {journal} {\bibinfo  {journal} {Physical Review D}\ }\textbf {\bibinfo
  {volume} {88}},\ \bibinfo {pages} {103501} (\bibinfo {year}
  {2013})}\BibitemShut {NoStop}%
\bibitem [{\citenamefont {Kanyolo}\ and\ \citenamefont
  {Masese}(2020)}]{kanyolo2020idealised}%
  \BibitemOpen
  \bibfield  {author} {\bibinfo {author} {\bibfnamefont {G.~M.}\ \bibnamefont
  {Kanyolo}}\ and\ \bibinfo {author} {\bibfnamefont {T.}~\bibnamefont
  {Masese}},\ }\bibfield  {title} {\bibinfo {title} {An idealised approach of
  geometry and topology to the diffusion of cations in honeycomb layered oxide
  frameworks},\ }\href@noop {} {\bibfield  {journal} {\bibinfo  {journal}
  {Scientific Reports}\ }\textbf {\bibinfo {volume} {10}} (\bibinfo {year}
  {2020})}\BibitemShut {NoStop}%
\bibitem [{\citenamefont {Milgrom}(1983)}]{milgrom1983modification}%
  \BibitemOpen
  \bibfield  {author} {\bibinfo {author} {\bibfnamefont {M.}~\bibnamefont
  {Milgrom}},\ }\bibfield  {title} {\bibinfo {title} {A modification of the
  newtonian dynamics as a possible alternative to the hidden mass hypothesis},\
  }\href@noop {} {\bibfield  {journal} {\bibinfo  {journal} {The Astrophysical
  Journal}\ }\textbf {\bibinfo {volume} {270}},\ \bibinfo {pages} {365}
  (\bibinfo {year} {1983})}\BibitemShut {NoStop}%
\bibitem [{Note5()}]{Note5}%
  \BibitemOpen
  \bibinfo {note} {Since $r = \protect \sqrt {x^2 + y^2 + z^2}$, $r_{\infty } =
  1/a_{0}$ implies that $z_{\infty } = 1/3a_{\infty }$, where the space is
  isotropic.}\BibitemShut {Stop}%
\bibitem [{\citenamefont {McGaugh}\ \emph {et~al.}(1995)\citenamefont
  {McGaugh}, \citenamefont {Bothun},\ and\ \citenamefont
  {Schombert}}]{mcgaugh1995galaxy}%
  \BibitemOpen
  \bibfield  {author} {\bibinfo {author} {\bibfnamefont {S.~S.}\ \bibnamefont
  {McGaugh}}, \bibinfo {author} {\bibfnamefont {G.~D.}\ \bibnamefont
  {Bothun}},\ and\ \bibinfo {author} {\bibfnamefont {J.~M.}\ \bibnamefont
  {Schombert}},\ }\bibfield  {title} {\bibinfo {title} {Galaxy selection and
  the surface brightness distribution},\ }\href@noop {} {\bibfield  {journal}
  {\bibinfo  {journal} {arXiv preprint astro-ph/9505062}\ } (\bibinfo {year}
  {1995})}\BibitemShut {NoStop}%
\bibitem [{Note6()}]{Note6}%
  \BibitemOpen
  \bibinfo {note} {{\protect \color {black} To simplify the expressions, one
  can re-scale $\phi \rightarrow \protect \bar {m}_{\protect \rm P}\phi $, and
  then make the choice, $\phi = M\DOTSI \intop \ilimits@ u_{\mu }dx^{\mu }$,
  which yields the geodesic equation when varied with respect to $x^{\mu
  }$.}}\BibitemShut {Stop}%
\bibitem [{\citenamefont {Buzano}\ \emph {et~al.}(2019)\citenamefont {Buzano}
  \emph {et~al.}}]{buzano2019higher}%
  \BibitemOpen
  \bibfield  {author} {\bibinfo {author} {\bibfnamefont {R.}~\bibnamefont
  {Buzano}} \emph {et~al.},\ }\bibfield  {title} {\bibinfo {title} {The
  higher-dimensional chern--gauss--bonnet formula for singular conformally flat
  manifolds},\ }\href@noop {} {\bibfield  {journal} {\bibinfo  {journal} {The
  Journal of Geometric Analysis}\ }\textbf {\bibinfo {volume} {29}},\ \bibinfo
  {pages} {1043} (\bibinfo {year} {2019})}\BibitemShut {NoStop}%
\bibitem [{\citenamefont {Lovelock}(1971)}]{lovelock1971einstein}%
  \BibitemOpen
  \bibfield  {author} {\bibinfo {author} {\bibfnamefont {D.}~\bibnamefont
  {Lovelock}},\ }\bibfield  {title} {\bibinfo {title} {The einstein tensor and
  its generalizations},\ }\href@noop {} {\bibfield  {journal} {\bibinfo
  {journal} {Journal of Mathematical Physics}\ }\textbf {\bibinfo {volume}
  {12}},\ \bibinfo {pages} {498} (\bibinfo {year} {1971})}\BibitemShut
  {NoStop}%
\bibitem [{\citenamefont {Kanyolo}\ and\ \citenamefont
  {Masese}(2021)}]{Kanyolo2021}%
  \BibitemOpen
  \bibfield  {author} {\bibinfo {author} {\bibfnamefont {G.~M.}\ \bibnamefont
  {Kanyolo}}\ and\ \bibinfo {author} {\bibfnamefont {T.}~\bibnamefont
  {Masese}},\ }\bibfield  {title} {\bibinfo {title} {Partition function for
  quantum gravity in 4 dimensions as a $1/\mathcal{N}$ expansion}} (\bibinfo
  {year} {2021}),\ \bibinfo {note} {hal-03335930}\BibitemShut {NoStop}%
\bibitem [{\citenamefont {Mak}\ and\ \citenamefont
  {Harko}(2013)}]{mak2013isotropic}%
  \BibitemOpen
  \bibfield  {author} {\bibinfo {author} {\bibfnamefont {M.}~\bibnamefont
  {Mak}}\ and\ \bibinfo {author} {\bibfnamefont {T.}~\bibnamefont {Harko}},\
  }\bibfield  {title} {\bibinfo {title} {Isotropic stars in general
  relativity},\ }\href@noop {} {\bibfield  {journal} {\bibinfo  {journal} {The
  European Physical Journal C}\ }\textbf {\bibinfo {volume} {73}},\ \bibinfo
  {pages} {2585} (\bibinfo {year} {2013})}\BibitemShut {NoStop}%
\bibitem [{\citenamefont {Thorne}\ \emph {et~al.}(2000)\citenamefont {Thorne},
  \citenamefont {Misner},\ and\ \citenamefont
  {Wheeler}}]{thorne2000gravitation}%
  \BibitemOpen
  \bibfield  {author} {\bibinfo {author} {\bibfnamefont {K.~S.}\ \bibnamefont
  {Thorne}}, \bibinfo {author} {\bibfnamefont {C.~W.}\ \bibnamefont {Misner}},\
  and\ \bibinfo {author} {\bibfnamefont {J.~A.}\ \bibnamefont {Wheeler}},\
  }\href@noop {} {\emph {\bibinfo {title} {Gravitation}}}\ (\bibinfo
  {publisher} {Freeman},\ \bibinfo {year} {2000})\BibitemShut {NoStop}%
\bibitem [{\citenamefont {Nariai}(1950)}]{nariai1950some}%
  \BibitemOpen
  \bibfield  {author} {\bibinfo {author} {\bibfnamefont {H.}~\bibnamefont
  {Nariai}},\ }\bibfield  {title} {\bibinfo {title} {On some static solutions
  of einstein's gravitational field equations in a spherically symmetric
  case},\ }\href@noop {} {\bibfield  {journal} {\bibinfo  {journal} {Scientific
  Reports of Tohoku University}\ }\textbf {\bibinfo {volume} {34}},\ \bibinfo
  {pages} {160} (\bibinfo {year} {1950})}\BibitemShut {NoStop}%
\bibitem [{\citenamefont {Nariai}(1999)}]{nariai1999new}%
  \BibitemOpen
  \bibfield  {author} {\bibinfo {author} {\bibfnamefont {H.}~\bibnamefont
  {Nariai}},\ }\bibfield  {title} {\bibinfo {title} {On a new cosmological
  solution of einstein's field equations of gravitation},\ }\href@noop {}
  {\bibfield  {journal} {\bibinfo  {journal} {General Relativity and
  Gravitation}\ }\textbf {\bibinfo {volume} {31}},\ \bibinfo {pages} {963}
  (\bibinfo {year} {1999})}\BibitemShut {NoStop}%
\end{thebibliography}%

\end{document}